\documentclass[aps,floatfix]{revtex4-1}
\usepackage{textcomp}
\usepackage{graphicx}
\graphicspath{{figure/}{..figure/}}
\usepackage{pstricks}    
\usepackage{amsmath,amssymb,amsfonts}
\usepackage{bm}
\usepackage{algorithmic}
\usepackage{array}
\usepackage[caption=false]{subfig}
\usepackage{cleveref}
\usepackage{color,soul}

\newcommand*{\DBtensor}[1]{\overline{\overline{#1}}}

\begin{document}
%
\title{Magneto-electric Uniaxial Metamaterials as Wide-angle, Polarization-insensitive Matching Layers}
\author{Yuchu He}
\affiliation{The Edward S. Rogers Sr. Department of Electrical and Computer Engineering, University of Toronto, Toronto, Ontario, Canada}%
\author{George V. Eleftheriades}%
\affiliation{The Edward S. Rogers Sr. Department of Electrical and Computer Engineering, University of Toronto, Toronto, Ontario, Canada}%

\begin{abstract}
{\color{black}
Antireflection or impedance matching is a topic that has been extensively researched by the optical and microwave communities over the past century and until today. However, due to the diverging wave impedances of TE ({\it s}) and TM ({\it p}) polarizations with increasing incident angle, it is impossible to achieve perfect matching simultaneously for both polarizations at varying incidence angles with a single conventional isotropic matching layer. To achieve polarization-insensitive matching at an arbitrary incident angle, we propose a magneto-electric uniaxial metamaterial layer (MEUML) that is inspired by the perfectly match layer (PML) concept in computational electromagnetics. Similar to the PML, the MEUML requires specific uniaxial permittivity and permeability tensors. However, to simultaneously control both the transversal and longitudinal material parameters is not an easy task. To date, a true PML has not been realized with metamaterials. In this paper, we employ a simple and yet special metamaterial unit cell to achieve such a control and synthesize a physical MEUML. The unit cell comprises two parallel metallic rings separated by a holey substrate. The transversal electric and magnetic dipole moments, and the longitudinal capacitive and diamagnetic coupling between the rings are judiciously controlled to achieve the required permittivity and permeability tensors. To aid the MEUML synthesis, we also introduce a technique that can extract the material parameter tensors at any incident angle. We first demonstrate this concept by achieving polarization-insensitive matching of a high-index substrate at $45^\circ$ with a single sub-wavelength thick MEUML. We further adapt the concept to the microwave regime by developing a MEUML based radome. Exceptional matching performance was obtained both in the simulations and measurements. The reflectance remains below 5\% from normal incidence ($0^\circ$) to near grazing angle ($85^\circ$) for both polarizations and over a wide bandwidth. With this unprecedented control of the material parameter tensors, the MEUML concept not only can be applied to impedance matching, but also can be utilized in many exotic applications that require extreme control of the material properties. 
}
\end{abstract}
\maketitle


%

\section{Introduction}
When an electromagnetic wave impinges on an interface between two media with different refractive indices, the wave will partially reflect and transmit due to the mismatch of the wave impedances between the two media. In many optical or microwave applications, such reflection is undesirable when maximum transmission is required. This has led to the development of antireflection theory and a plethora of antireflection structures. The simplest antireflection structure is the quarter-wave transformer (QWT) \cite{Cheng} that consists of a single dielectric layer with a refractive index of $\sqrt{n_s}$, where $n_s$ is the index of the substrate to be matched. Despite its simplicity, the QWT is often limited in applications due to the strict index and thickness requirements. In addition, the QWT only achieves perfect matching at normal incidence. Reflectance quickly rises as the incident angle becomes more oblique. More complex structures, such as multi-layer coatings \cite{Macleod_Thin_2001, Dobrowolski_Toward_2002} and moth-eye type graded-index (GRIN) coatings \cite{CLAAM_Reduction_1973} can be used to extend the angular and frequency ranges, but they are usually thick and complex for fabrication. In addition, these thick coatings are usually impractical at longer wavelengths due to their significant thickness and weight. A good overview of various antireflection coatings can be found in \cite{Raut_Anti_2011}. 

So far, all these coatings which are based on  conventional antireflection theory usually utilize isotropic materials with positive $\varepsilon$ and unity $\mu$, corresponding to common natural materials. With the developments in metamaterials \cite{Pendry_Magnetism_1999,Pendry_Negative_2000,Smith_Negative_2000}, conventional material constraints can be circumvented. Such artificially engineered structures may possess a wide range of $\varepsilon$ and $\mu$ values, including negative ones \cite{Smith_Composite_2000,Smith_Electromagnetic_2003,Shelby77,PhysRevLett.95.137404}. This has led to the development of metamaterial based reflection-less absorbers \cite{PhysRevLett.100.207402,PhysRevLett.107.045901,Na_liu_IR_absorber} and low-loss antireflection layers \cite{Metamaterial_AR_PRL,Metamaterial_AR_SciRep}. Even though these metamaterial-based antireflection layers are more compact, they work well only in a narrow angular range near normal incidence.

The issue of minimizing reflections at large oblique angles of incidence was previously addressed in \cite{7936430}. A matching theory based on anisotropic metamaterials was developed for matching an arbitrary substrate at an arbitrary angle of incidence, including extreme ones. By employing anisotropic metamaterial layers, the authors demonstrated perfect matching at $88^\circ$ for either TE or TM polarization. At such an extreme angle, matching with conventional methods becomes difficult if not impossible. In addition, the resulting anisotropic layer is sub-wavelength thick, making it much more practical than the $120\lambda_0$ thick GRIN layer that was numerically investigated in \cite{Dobrowolski_Toward_2002}.
Nevertheless, the shortcoming of this anisotropic matching layer is that the layer has to be designed to match either TE or TM polarizations, but not both. This is because the wave impedances for the two polarizations diverge, as the incidence angle increases. Thus, it is difficult to design matching structures to match for both polarizations. Conventional coatings are often prioritized to match one of the two polarizations. 


{\color{black} Polarization-insensitive matching was hinted by the perfectly matched layer (PML) \cite{Gedney_An_1996} concept in the field of computational electromagnetics. The PML is capable of absorbing an incident wave without reflection, regardless of the incident angle and polarization. To achieve this perfect matching, the PML has identical and lossy $\varepsilon$ and $\mu$ tensors in uniaxial form. Inspired by the PML, in this paper, we present a polarization-insensitive metamaterial matching layer with uniaxial permittivity and permeability tensors. However, we would like to point out that even though the MEUML has some resemblance to the PML, they are far from being equivalent. The most crucial difference is that the PML is an absorbing medium whereas the MEUML is not. As a result, the MEUML has to satisfy one more boundary condition than the PML. Since the PML is absorbing, minimal amount of power reaches the back surface of the PML, and the boundary condition at the back surface of the PML is not relevant. In fact, one can use a perfect electric conductor (PEC) as the boundary without introducing large reflections. In comparison, the boundary condition at the back surface of the MEUML (the MEUML/substrate interface) is crucial. Different amounts of reflections can arise at the interface for different substrates and at different incident angles. To achieve perfect matching, the required $\varepsilon$ and $\mu$ tensors depend on the substrate and incident angle, and they are usually not identical. Thus, the MEUML is a more constraint, non-lossy version of the PML.
For non-absorbing structures, an all angle transmitting and polarization-insensitive Huygens' metasurface was proposed in \cite{7304824}. The authors derived analytical angularly dependent surface impedance tensors, which enable the metasurface to transmit an incident plane wave at any angle without reflection. However, this type of metasurface only works for the special case of transmittance from free space to free space. It would not work if one side is changed an arbitrary medium. In comparison, the MEUML is more general and can be easily extended to matching between two arbitrary media.

So far, physical implementation of these absorbing and non-absorbing structures are far from perfect. For the non-absorbing Huygens' metasurface in \cite{7304824}, the authors proposed to implement such a metasurface as a thin layer with a finite thickness and with the surface impedances approximated with tangential and longitudinal permittivities and permeabilities. However, the authors of \cite{7304824} did not provide a physical structure to realize these material parameter tensors. On the other hand, for absorbing structures, a true PML has not been physically realized with metamaterials. The implementations are always an approximate form. Previously reported metamaterials-based absorbers \cite{cui2014plasmonic,PhysRevLett.100.207402,ye2013ultrawideband,ye2012towards,PhysRevB.79.045131,Na_liu_IR_absorber,he2013infrared} do not work exactly as the theoretical PML. Absorbers in \cite{PhysRevLett.100.207402,ye2013ultrawideband,ye2012towards} are resonance based; thus, the absorption can be narrow band, angle sensitive, and polarization sensitive. Absorbers in \cite{ye2013ultrawideband,ye2012towards,PhysRevB.79.045131,Na_liu_IR_absorber} only demonstrated control of the transversal parameters while the control of the longitudinal parameters is absent. According to the PML theory, the longitudinal parameters should be the inverse of the transversal parameters. For the absorber in \cite{he2013infrared}, the longitudinal permittivity can be tuned for an absorber based on a nanowire array, but the control on the magnetic parts is absent. Comparing to the PML with identical $\varepsilon$ and $\mu$ tensors, realizing a MEUML with non-identical $\varepsilon$ and $\mu$ tensors is a much more daunting task.

In this paper, we are proposing a novel metamaterial unit cell to realize the more complex MEUML. By carefully controlling the transversal and longitudinal electric and magnetic coupling in the unit cell, we can simultaneously tune the uniaxial permittivity and permeability tensors to the desired values. The synthesis process is aided by a parameter extraction technique we provide in the Appendix, which can be used to accurately extract all the layer parameter values at any incident angle. For demonstration purposes, the MEUML is designed to match to a high-index substrate at $45^\circ$ incidence and at a design frequency of 10~GHz. Around -30~dB reflection levels are observed for both polarizations. 
The proposed MEUML is a significant extension to our previous work in \cite{7936430}, in which the anisotropic matching layer works for either TE or TM polarization, but not both; and the matching layers are designed for a specific orientation in the azimuthal plane ($y-z$ plane). In comparison, the MEUML not only matches for both polarizations simultaneously, but also for all azimuthal orientations. }

To further demonstrate the new matching possibilities enabled by the MEUML, we adapt the MEUML to impedance matching in the microwave regime, which has a slightly different design philosophy than in the optical regime. For optical applications, the matching layer or the coating is usually designed to match from air to an optical substrate. Since typical optical substrates are optically thick (thousands of wavelengths), they can be assumed to be semi-infinite.  Thus, when designing the matching layer, it can be assumed that the matching layer is sandwiched between a semi-infinite air region and a semi-infinite substrate. In contrast, the assumption of a semi-finite substrate is typically not valid in the microwave regime since the wavelength is much longer. Hence, the matching layer is designed to match from air to air. This kind of matching layer is essentially a radome. By using the transfer matrix method (TMM) \cite{Collin_Field_1960}, a MEUML based radome is designed and optimized. The reflection levels of the realized $\lambda_0/3$ thick radome remain below -14 dB (4\% power reflection) from $0^\circ$ at normal incidence to $85^\circ$ near the grazing angle for both polarizations. Comparing to radomes based on half-wavelength dielectrics \cite{lo2013antenna}, sandwich structures with high/low/high index dielectrics \cite{cady1948radar}, metamaterial based frequency selective surfaces (FSS) \cite{1140896,pelletti2013three,6931715,8072179}, multilayer dielectrics \cite{kedar2006parametric}, and inhomogeneous dielectrics \cite{6253228}, which all have reflections increasing rapidly beyond $60^\circ$ or $70^\circ$, the proposed MEUML radome is superior in angular stability and polarization-insensitivity. To our best knowledge, this is the widest angular range that has been achieved for maintaining less than 5\% power reflection for both polarizations and over a wide bandwidth.



\section{MEUML Matching Concept}\label{sec:MEUML Matching Concept}
\subsection{Matching Theory}
\begin{figure}[h!]
\centering
\includegraphics[width=0.35\columnwidth]{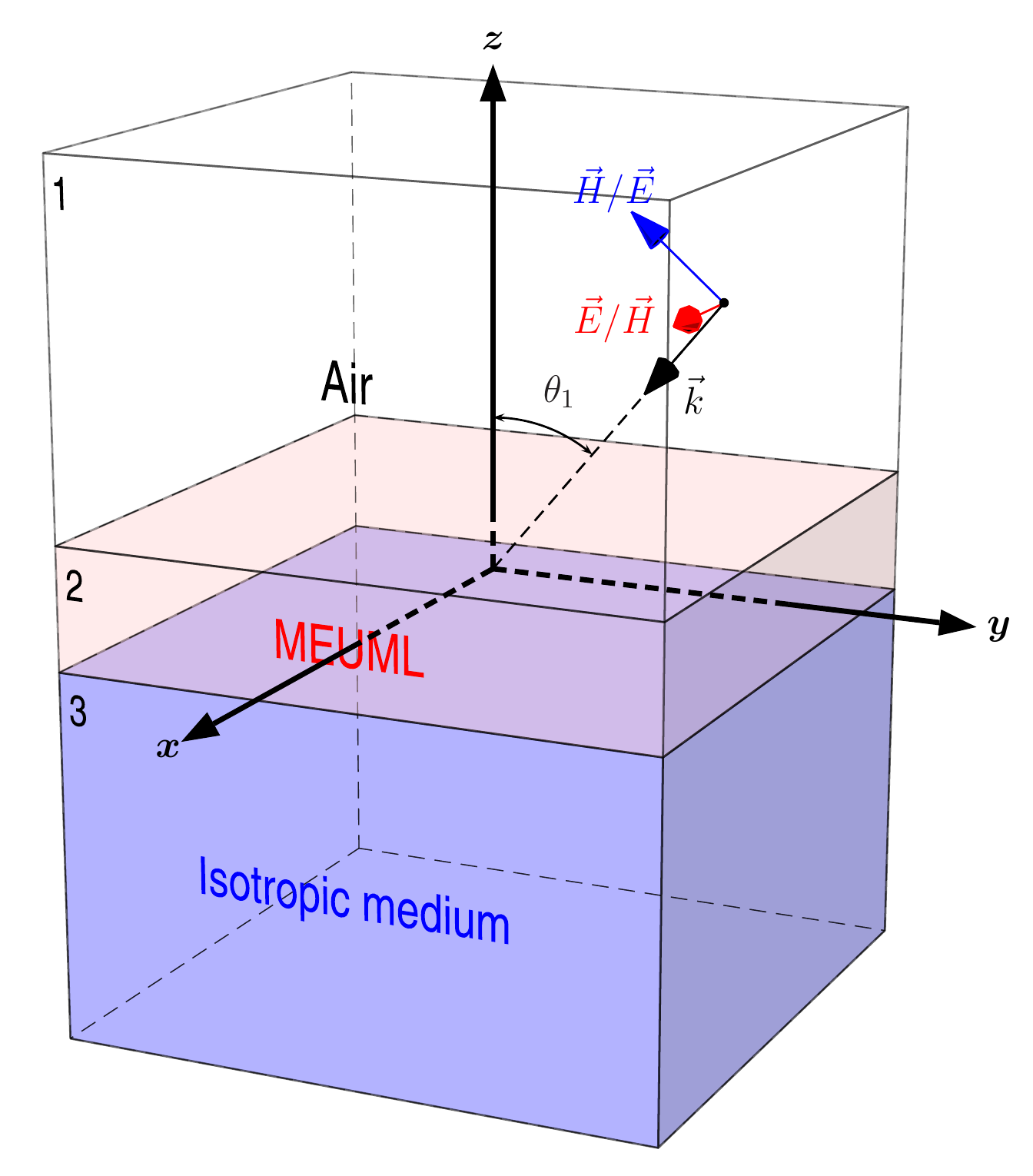}
\caption{A plane wave impinges on the MEUML at $\theta_1$ with an arbitrary polarization. The MEUML has thickness $d$ and sits on a semi-infinite isotropic medium that is to be matched. }
\label{fig:Incident_planewave}
\end{figure}

As shown in Fig. \ref{fig:Incident_planewave}, a plane wave with TE or TM polarization from air (medium 1) impinges on the MEUML (medium 2) at an angle of $\theta_1$. The MEUML sits on an isotropic non-magnetic semi-infinite substrate (medium 3) with an arbitrary permittivity $\varepsilon_3$.
The permittivity and permeability tensors of the uniaxial layer assume the following form
\begin{equation}
			 \DBtensor{\varepsilon}_{2}=
		\begin{bmatrix}
	 \varepsilon_{2t}  &  0  & 0 \\ 
		 0 & \varepsilon_{2t} & 0\\
		0 & 0 &	\varepsilon_{2n}
			\end{bmatrix}
\label{Eq:tensor_epsilon}
\end{equation}
\begin{equation}
			 \DBtensor{\mu}_{2}=
		\begin{bmatrix}
	 \mu_{2t}  &  0  & 0 \\ 
		 0 & \mu_{2t} & 0\\
		0 & 0 &	\mu_{2n}
			\end{bmatrix}
\label{Eq:tensor_mu}
\end{equation}
where subscript $t$ denotes the tangential material parameters in the  $x$ and $y$ direction,  and subscript $n$ denotes the longitudinal material parameters along the $z$ direction. In addition, all the matrix elements are assumed to be real. The total reflection in air can be calculated from \eqref{Eq:total_r}
\begin{equation}
r^\mathrm{TM/TE}=\frac{r_{12}^\mathrm{TM/TE}+r_{23}^\mathrm{TM/TE}e^{-i2\phi^\mathrm{TM/TE}}}{1+r_{12}^\mathrm{TM/TE}r_{23}^\mathrm{TM/TE}e^{-i2\phi^\mathrm{TM/TE}}}
\label{Eq:total_r}
\end{equation}
where $r_{ij}^\mathrm{TM/TE}$ is the Fresnel reflection coefficient at the $i$, $j$ media interface with the incidence from medium $i$; the phase $\phi^\mathrm{TM/TE}$ is the total phase delay accumulated in the uniaxial layer along the normal of the layer surface (z-axis). 
By equating \eqref{Eq:total_r} to zero, the permittivity and permeability tensors in \eqref{Eq:tensor_epsilon} and \eqref{Eq:tensor_mu} can be solved exactly (See Appendix~\ref{sec:MEUML_parameter_derivation}), 
\begin{equation}
\varepsilon_{2t}=\frac{\lambda_0}{4d}\frac{\sqrt{\varepsilon_3}}{\sqrt{\cos\theta_1\sqrt{\varepsilon_3-\sin^2\theta_1}}}
\label{Eq:eps_2t}
\end{equation}
\begin{equation}
\mu_{2t}=\frac{\lambda_0}{4d}\frac{1}{\sqrt{\cos\theta_1\sqrt{\varepsilon_3-\sin^2\theta_1}}}
\label{Eq:mu_2t}
\end{equation}
\begin{equation}
\varepsilon_{2n}=\frac{4d}{\lambda_0}\frac{\sqrt{\varepsilon_3}\sin^2\theta_1\sqrt{\cos\theta_1\sqrt{\varepsilon_3-\sin^2\theta_1}}}{\sqrt{\varepsilon_3}-\cos\theta_1\sqrt{\varepsilon_3-\sin^2\theta_1}}
\label{Eq:eps_2n}
\end{equation}
\begin{equation}
\mu_{2n}=\frac{4d}{\lambda_0}\frac{\sin^2\theta_1\sqrt{\cos\theta_1\sqrt{\varepsilon_3-\sin^2\theta_1}}}{\sqrt{\varepsilon_3}-\cos\theta_1\sqrt{\varepsilon_3-\sin^2\theta_1}}
\label{Eq:mu_2n}
\end{equation}
where $\lambda_0$ is the free-space wavelength and $d$ is the MEUML thickness. Using \eqref{Eq:eps_2t}-\eqref{Eq:mu_2n}, we can calculate the required MEUML parameters to achieve perfect matching for a particular substrate at a particular incident angle for both TE and TM polarizations. {\color{black} It is interesting to note that if we substitute $\varepsilon=1$ in to \eqref{Eq:eps_2t}-\eqref{Eq:mu_2n}, we arrive to the same conditions ((7) and (8))in \cite{7304824}. Thus, our formulation is more general and rigorous and it can be easily extend to the case of arbitrary media on both sides of the layer.}

\subsection{MEUML unit cell design}\label{sec:MEUML_design}
\begin{figure}[h!] 
    \centering
  \subfloat[]{%
       \includegraphics[height=6.7cm]{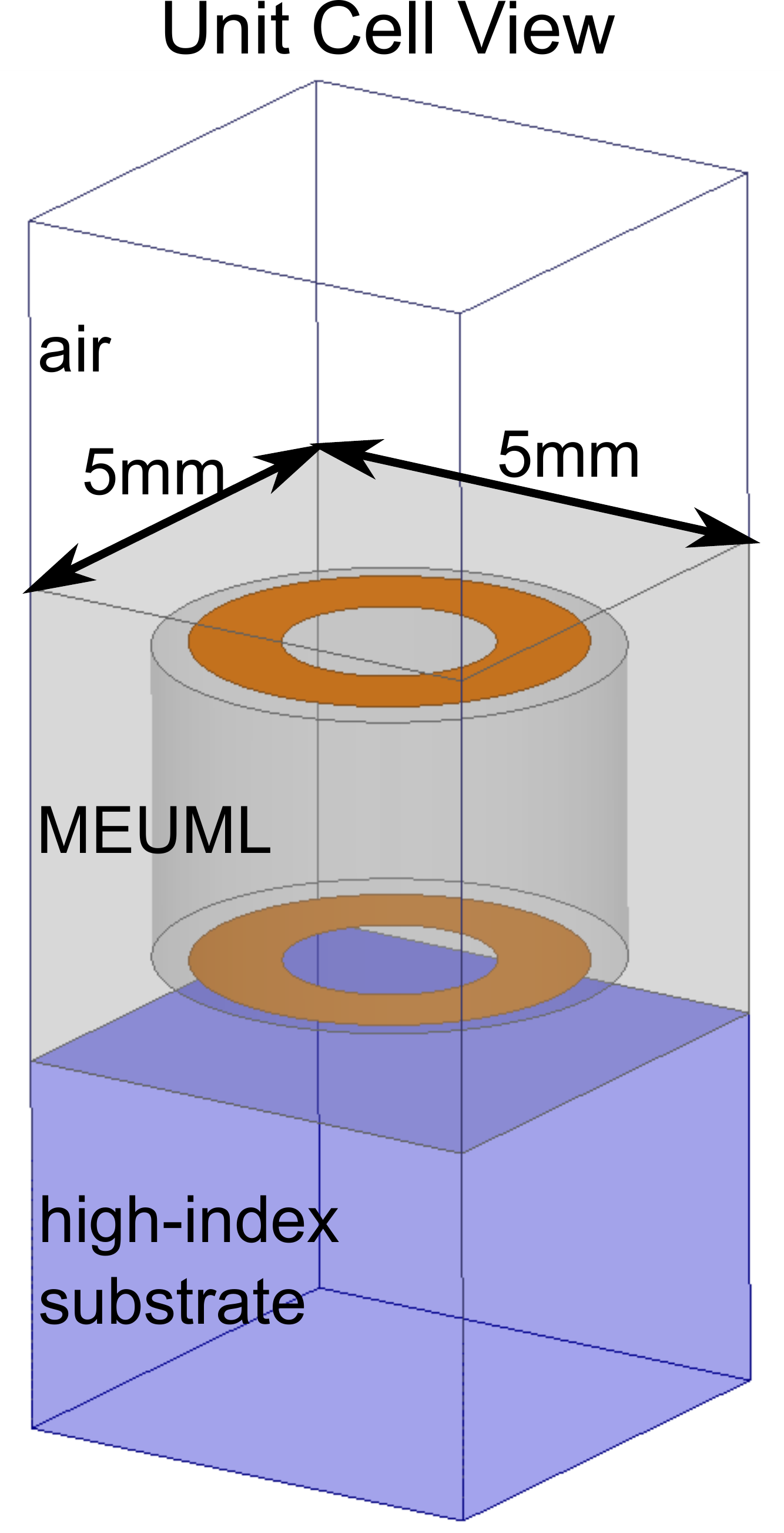}}
    \label{subfig:Unitcell_view}
  \subfloat[]{%
        \includegraphics[height=6.7cm]{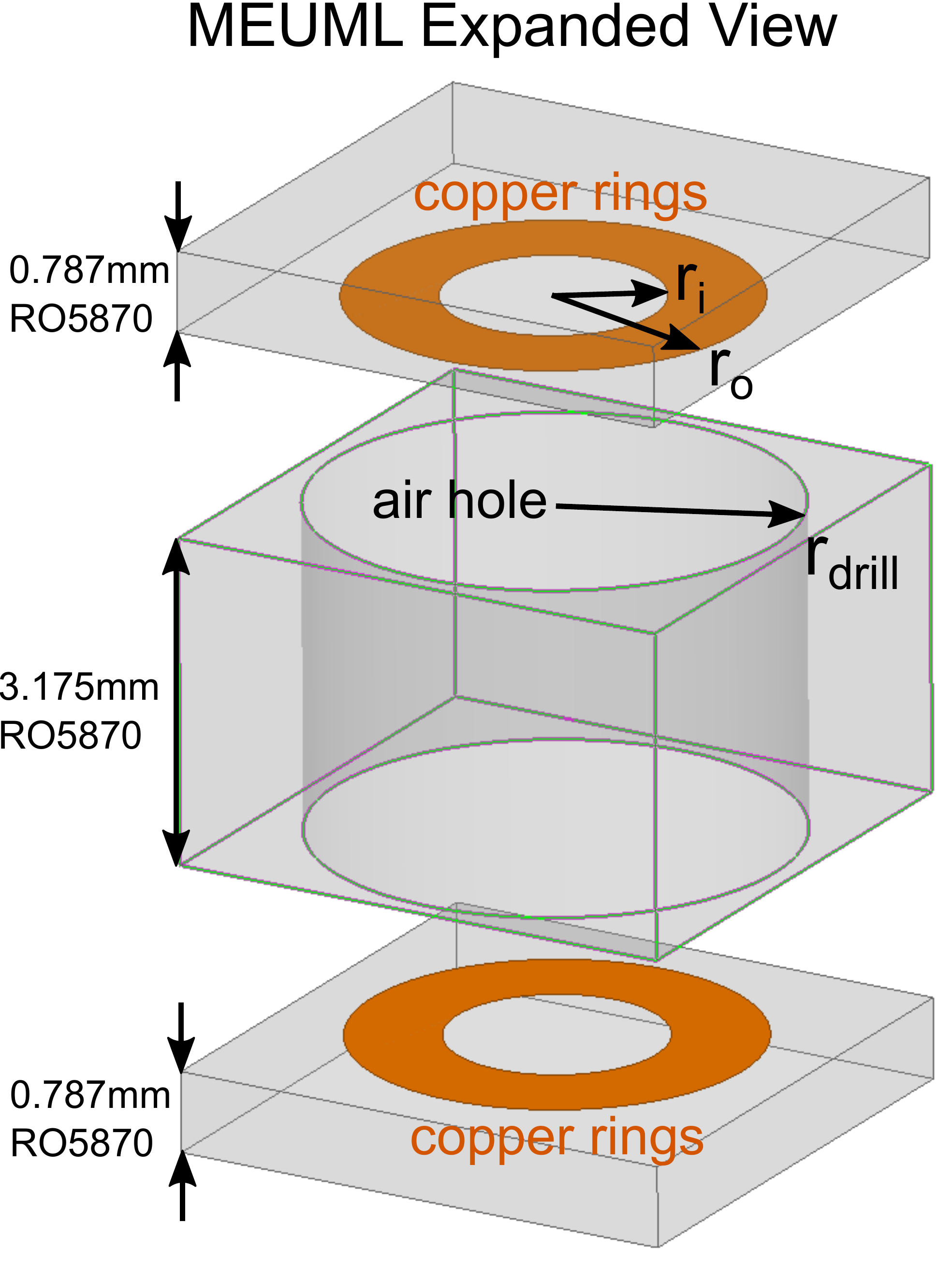}}
    \label{subfig:MEUML_exploded_view}
  \caption{{Unit cell design of the MEUML.} (a) In the unit cell view, the MEUML is sandwiched between a semi-infinite air region and a semi-infinite substrate. The host medium of the MEUML is a Rogers RO5870 ($\varepsilon_r=2.3$) slab with a total thickness of 4.75mm. 
(b) The physical realization of the MEUML. The 4.75mm slab is realized by bonding a 3.175~mm substrate to two 0.787~mm ones as shown in the exploded view. Two copper rings are patterned on the inner surfaces of the 0.787~mm substrates and an air hole is drilled through the 3.175~mm one. The geometries of the final structure are: $r_i=1.04$~mm, $r_o=1.94$~mm, and $r_{drill}=2.3$~mm.}
  \label{fig:Unitcell_design} 
\end{figure}

\begin{figure}[htb!] 
    \centering
  \subfloat[]{%
       \includegraphics[height=4.8cm]{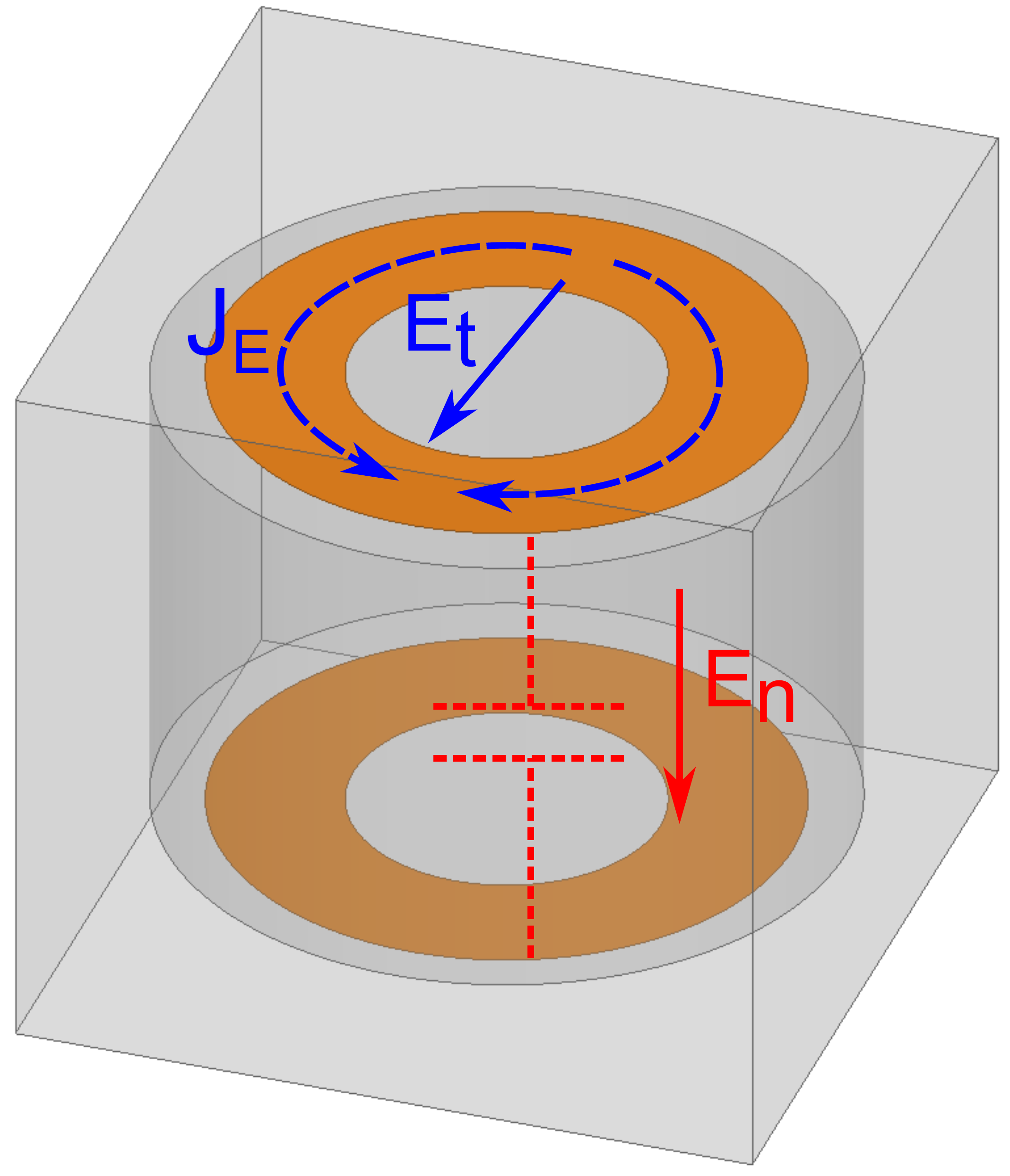}}
    \label{subfig:ring_mechanism_E_field}\quad
  \subfloat[]{%
        \includegraphics[height=4.8cm]{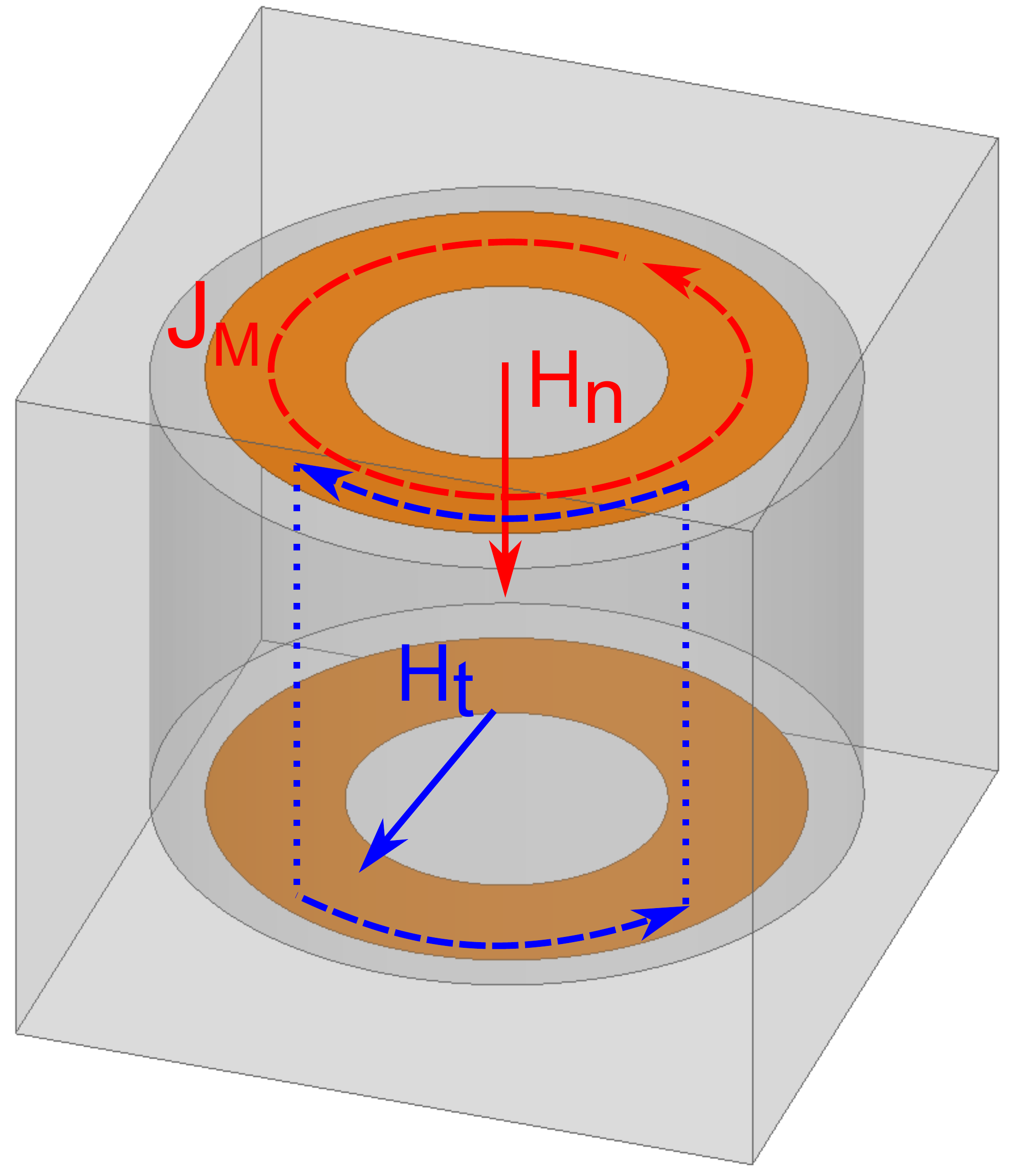}}
    \label{subfig:ring_mechanism_H_field}
  \caption{{ The working mechanism of the MEUML.} (a) The electric field of a TM-polarized wave at an oblique incidence can be decomposed into tangential and longitudinal components. The tangential electrical field induces dipolar currents flowing on the rings, which result to a capacitive loading and lead to an increase in $\varepsilon_{2t}$. The top and bottom rings can be treated as a capacitor; thus, $\varepsilon_{2n}$ can be tuned by controlling the spacing and the area of the rings. (b) The tangential magnetic field from the TE polarized component induces counter-flowing currents on the top and the bottom rings, which form a partial current loop. This induces a weak magnetic effect and can be used to tune $\mu_{2t}$ slightly from unity. On the other hand, the longitudinal magnetic field will induce current loops flowing around the rings. According to Lenz's law, the induced currents generate an opposing magnetic field to the incident one. Due to this diamagnetic effect, $\mu_{2n}$ is lowered. Since the required $\varepsilon_{2n}$ of 1.5 is lower than the host substrate permittivity, an air hole is used to lower the effective host permittivity and this in turn lowers the $\varepsilon_{2n}$ to the desired value.}
  \label{fig:ring_mechanism} 
\end{figure}

To demonstrate the matching concept, a MEUML is designed to match to a high-index substrate ($\varepsilon_r=10.2$) at $45^\circ$. The physical MEUML can be synthesized according to the iterative design procedure in Appendix.~\ref{sec:MEUML_synthesis}.
The final unit cell design is shown in Fig.~\ref{fig:Unitcell_design}. The MEUML is 4.75mm thick and the required material parameters are: $\varepsilon_{2t}=3.40$, $\varepsilon_{2n}=1.51$, $\mu_{2t}=1.06$, and $\mu_{2n}=0.62$. 
To achieve those parameter values, we propose to use two copper rings lying in the $x-y$ plane that are separated by an air hole. The working principle of the proposed unit cell can be explained with the help of Fig.~\ref{fig:ring_mechanism}. 
The electric and magnetic fields of the incident TE and TM polarizations can be decomposed into their tangential and longitudinal components. 
The tangential electric field will induce current $J_{E}$ flowing on both rings. The induced currents are dipolar in nature and will produce a capacitive loading as in conventional artificial dielectrics \cite{Collin_Field_1960}, which results to an increase in the effective tangential permittivity. This enable us to increase the effective $\varepsilon_{2t}$ from the host medium permittivity of 2.3 to the required permittivity of 3.4. 
From the longitudinal electric field point of view, the top and bottom rings constitute a capacitor and produce a capacitive loading. By changing the outer and inner ring radii, and the ring spacing, $\varepsilon_{2n}$ can be tuned. 
The longitudinal magnetic field will induce currents flowing around the rings \cite{Liu:08}. Since the rings are electrically small, they are diamagnetic according to Lenz's law. As a result, the effective longitudinal permeability is reduced due to this diamagnetic effect. By controlling the size of the rings, $\mu_{2n}$ can be tuned. 
The tangential magnetic field induces counter flowing currents on the top and bottom rings \cite{Liu:08,PhysRevX.4.041042}, which form a partial current loop. This partial current loop can  induce a weak magnetic response and causes a change in the effective tangential permeability. The effect generally gets weaker as the rings are spaced further apart. Thus, by controlling the spacing between the rings, $\mu_{2t}$ can be tuned slightly from unity.
Lastly, notice that the required $\varepsilon_{2n}$ is 1.51, which is lower than the substrate permittivity of 2.3. Due to the capacitive coupling between the top and the bottom rings, $\varepsilon_{2n}$ is always larger than the host medium permittivity. Thus, a hole is drilled into the substrate to dilute the host medium permittivity, which in turn lowers $\varepsilon_{2n}$. Varying the air hole diameter has minimal impact on the permeabilities, but it does change $\varepsilon_{2t}$. Thus, the copper ring sizes have to be adjusted accordingly. 
In general, the four material parameters are coupled and varying one geometric parameter can change all four material parameters simultaneously. To find optimal geometries, extensive parametric sweeps and optimizations were performed.

{\color{black} The extracted permittivities and permeabilities for the proposed unit cell are plotted against the incident angle in Fig.~\ref{fig:extracted_layer_parameters}. The corresponding parameter extraction technique is described in the Appendix \ref{sec:MEUML_parameter_extraction}.  Table~\ref{tab:layer_parameter_comparison} compares the extracted values and the required theoretical values for perfect matching at $45^\circ$. The extracted parameter values are in a good agreement with the theoretical ones.
Fig.~\ref{fig:TandRandloss} shows the simulated reflection, transmission, and loss of the proposed unit cell. Minimum reflections for both TE and TM polarizations are observed near $45^\circ$ as expected. Due to the low-loss nature of the unit cell design, the transmission is near unity at the design angle. Since the extracted material parameters deviate slightly from the theoretical values, the minimum reflection for TM polarization is at $40^\circ$. We would like to point out that in this unit cell design, the spacing between the rings is constrained to a standard thickness of 3.175mm. The results can be further improved if one is not limited to standard substrate thicknesses. In addition, the basic optimization routines in the commercial simulator \cite{HFSS} often do not output optimal results for this complexly coupled structure. A more sophisticated optimization routine can be employed but it is out of the scope of this paper. }

\begin{figure}[htb!] 
    \centering
  \subfloat[]{%
       \includegraphics[width=0.3\columnwidth]{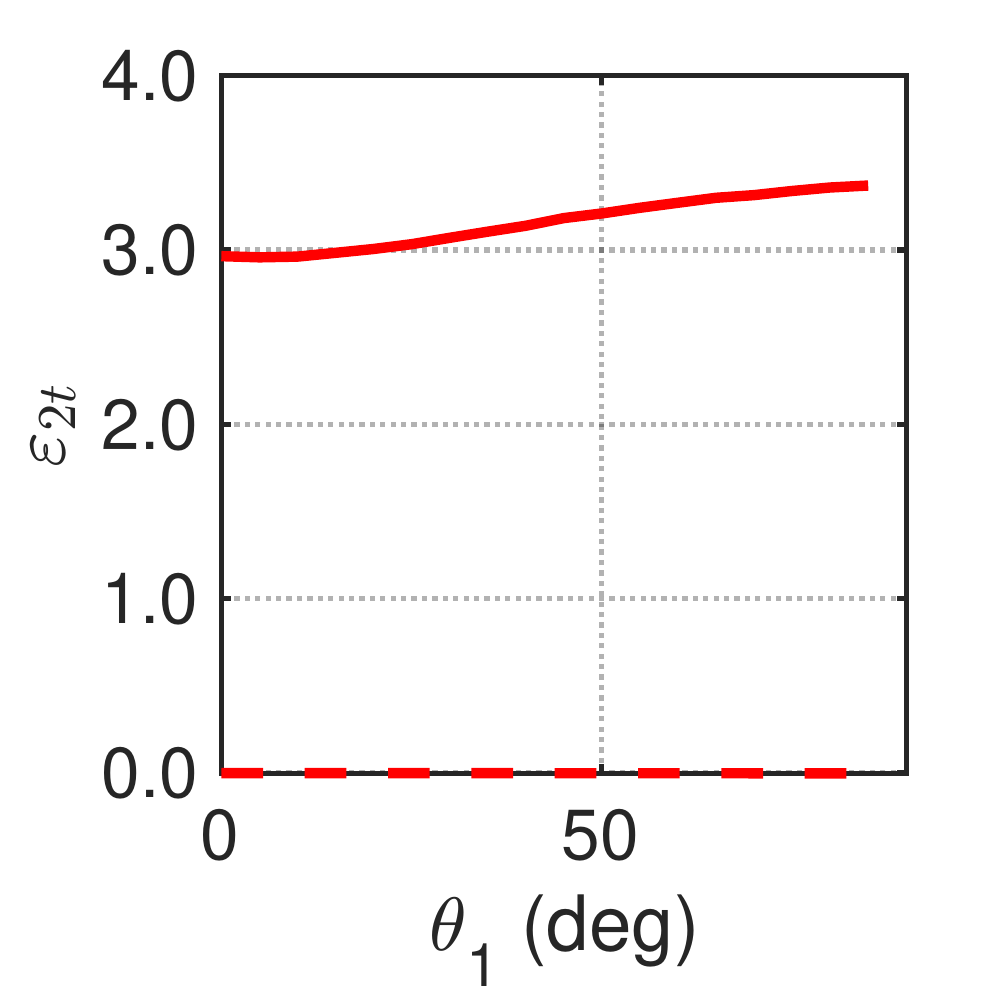}}
    \label{subfig:extracted_layer_parameters_eps_2t}
  \subfloat[]{%
        \includegraphics[width=0.4\columnwidth]{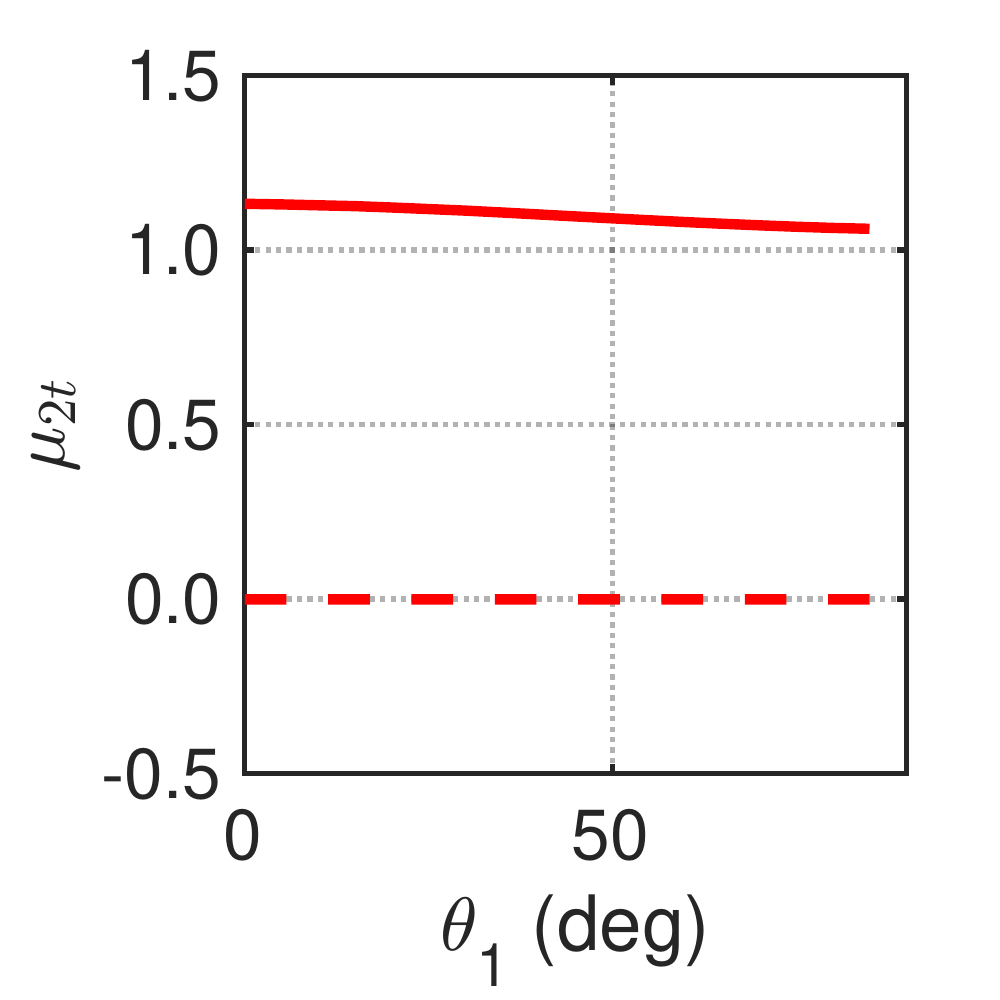}}		
    \label{subfig:extracted_layer_parameters_mu_2t}
		
	\subfloat[]{%
       \includegraphics[width=0.4\columnwidth]{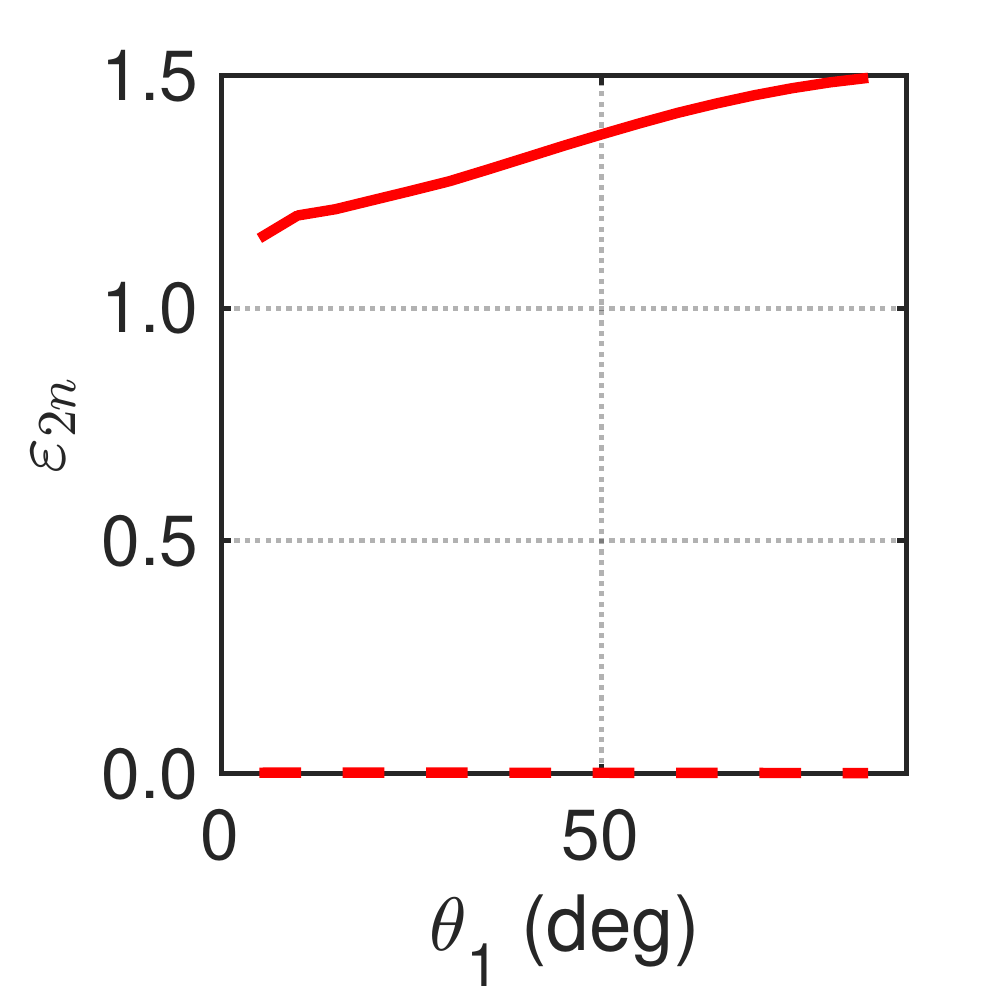}}
    \label{subfig:extracted_layer_parameters_eps_2tn}
  \subfloat[]{%
        \includegraphics[width=0.4\columnwidth]{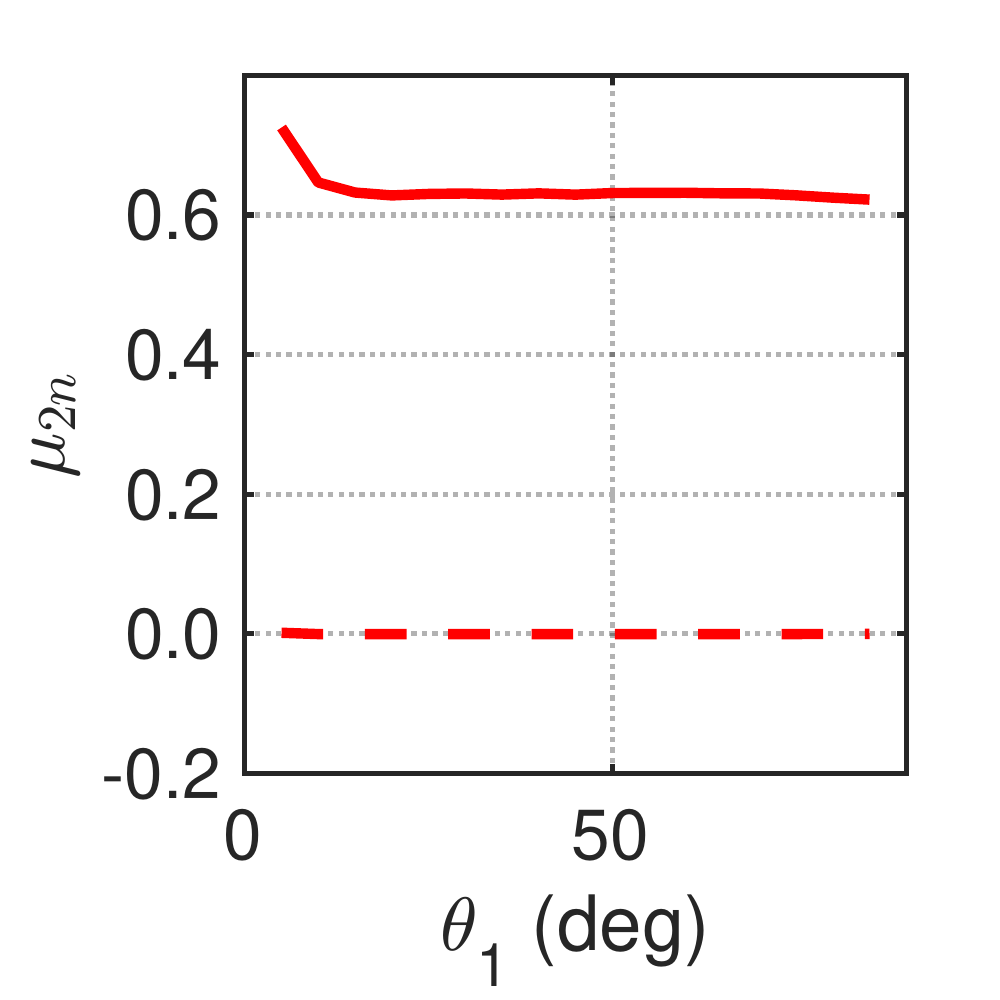}}		
    \label{subfig:extracted_layer_parameters_mu_2n}
  \caption{{ Extracted MEUML parameters as a function of the incident angle $\theta_1$.} The real parts (solid lines) and the imaginary parts (dashed lines) are shown for (a) $\varepsilon_{2t}$, (b) $\mu_{2t}$, (c) $\varepsilon_{2n}$, and (d) $\mu_{2n}$. The imaginary parts of all four parameters are near zero, indicating the MEUML having a small loss. At  normal incidence, the electromagnetic wave is independent from the longitudinal material parameters are undefined and cannot be extracted (See Appendix~\ref{sec:MEUML_parameter_extraction}).}
\label{fig:extracted_layer_parameters}
\end{figure}

\begin{table}[t!]
\caption{Theoretical vs. extracted MEUML parameters}
\centering
\begin{tabular}{p{50pt}>{\hangindent 0em}p{50pt}>{\hangindent 0em}p{50pt}}
\hline\hline
Parameters & Theoretical & Extracted \\
\hline
$\varepsilon_{2t}$ 	& $3.40$ 	& 3.20\\
$\mu_{2t}$ 					& $1.06$ 	& 1.09 \\
$\varepsilon_{2n}$ 	& $1.51$ 	& 1.40 \\
$\mu_{2n}$ 					& $0.47$ 	& 0.62 \\
\hline\hline
\end{tabular}
\label{tab:layer_parameter_comparison}
\end{table}

\begin{figure}[htb!] 
    \centering
  \subfloat[]{%
       \includegraphics[width=0.5\columnwidth]{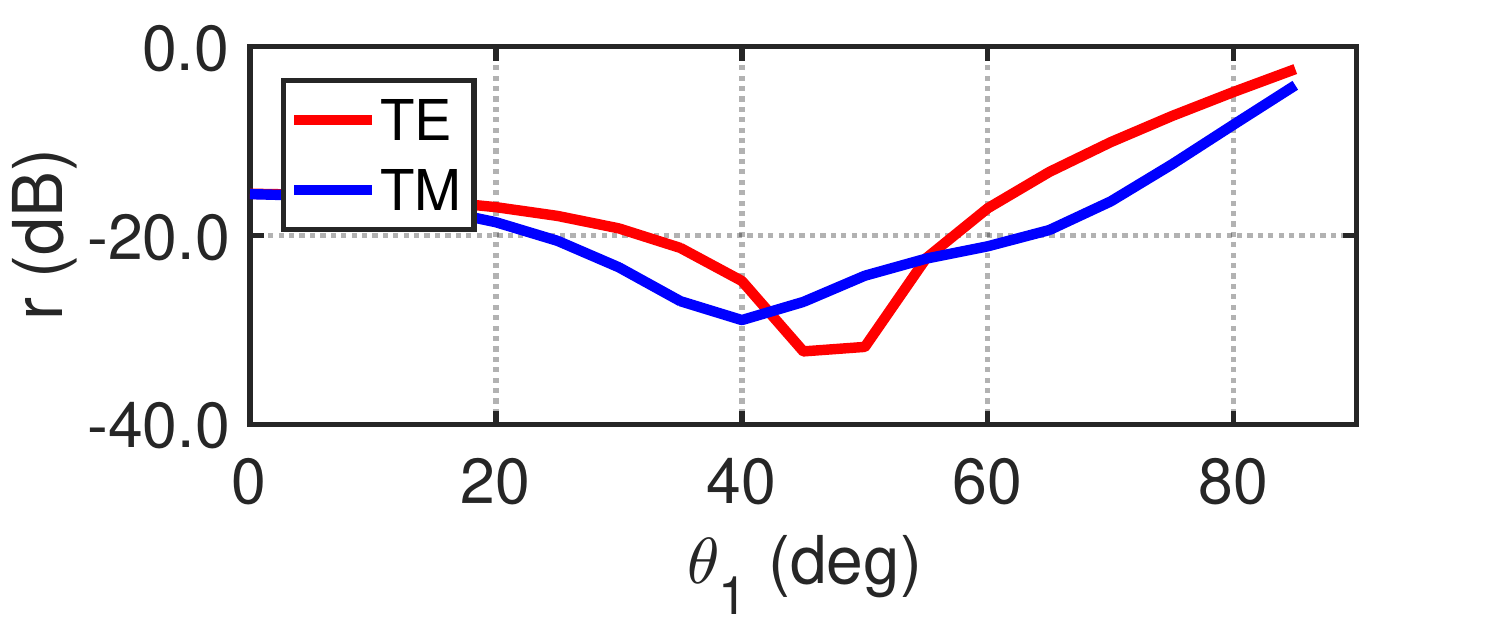}}
    \label{subfig:MEUML_r}
		
  \subfloat[]{%
        \includegraphics[width=0.5\columnwidth]{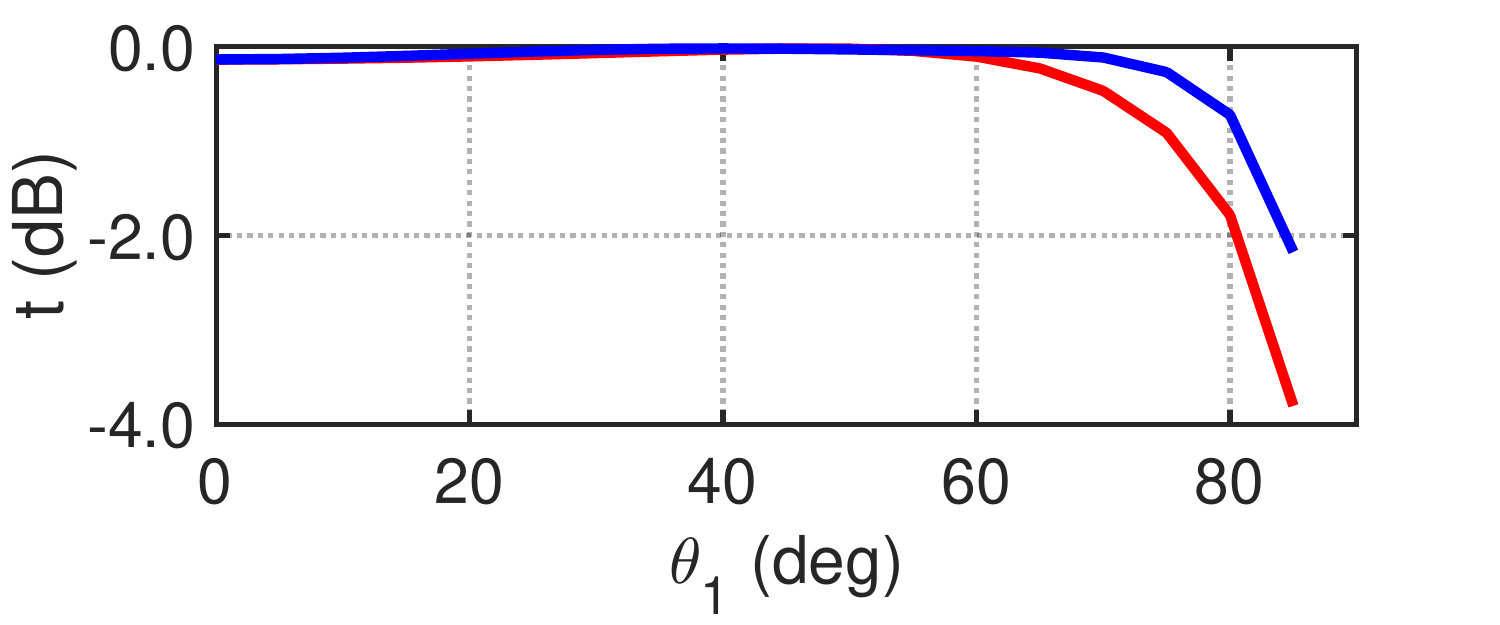}}		
    \label{subfig:MEUML_t}
		
	\subfloat[]{%
       \includegraphics[width=0.5\columnwidth]{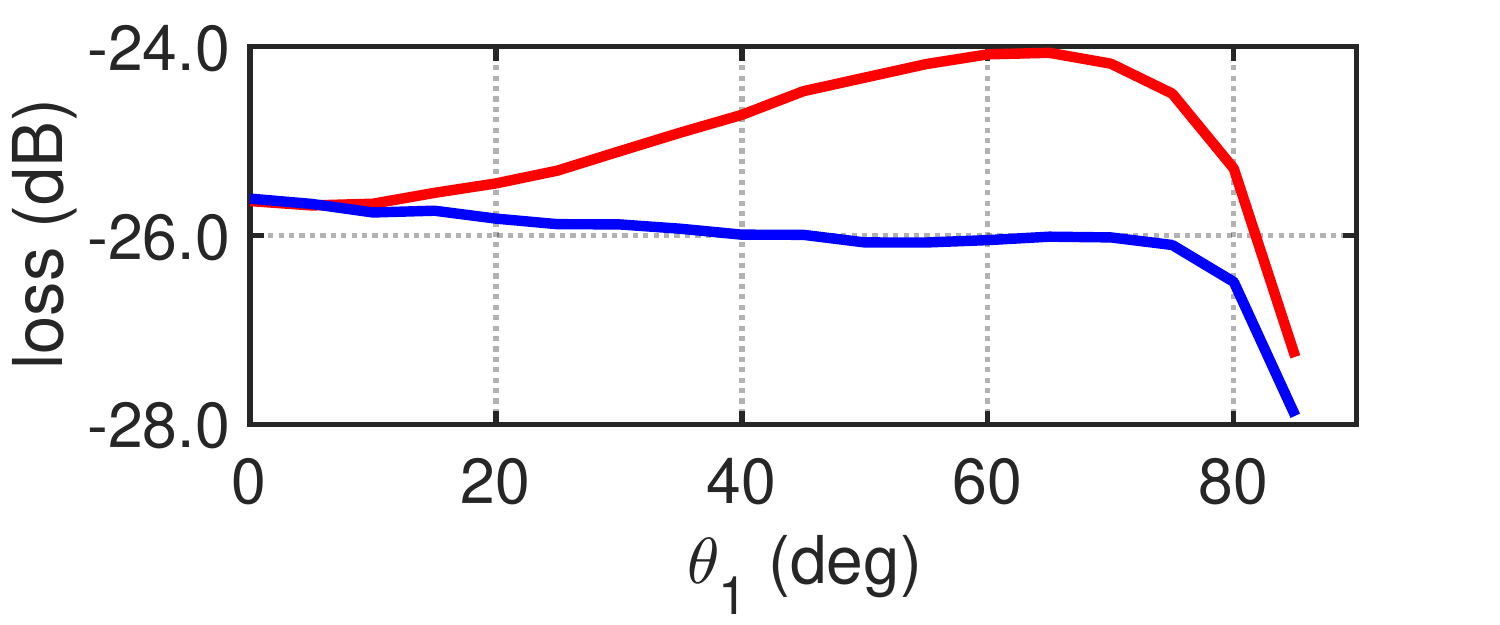}}
    \label{subfig:MEUML_loss}
		
  \caption{{The simulated responses of the MEUML.} (a) Reflection $r$, (b) transmission $t$, and (c) loss are plotted against $\theta_1$. Minimum reflections for both TE and TM polarizations are observed near $45^\circ$. Since the matching layer exhibits very small loss, the transmission is near unity at the design angle. }
\label{fig:TandRandloss}
\end{figure}

\begin{figure}[htb!]
    \centering
  \subfloat[]{%
       \includegraphics[width=0.3\linewidth]{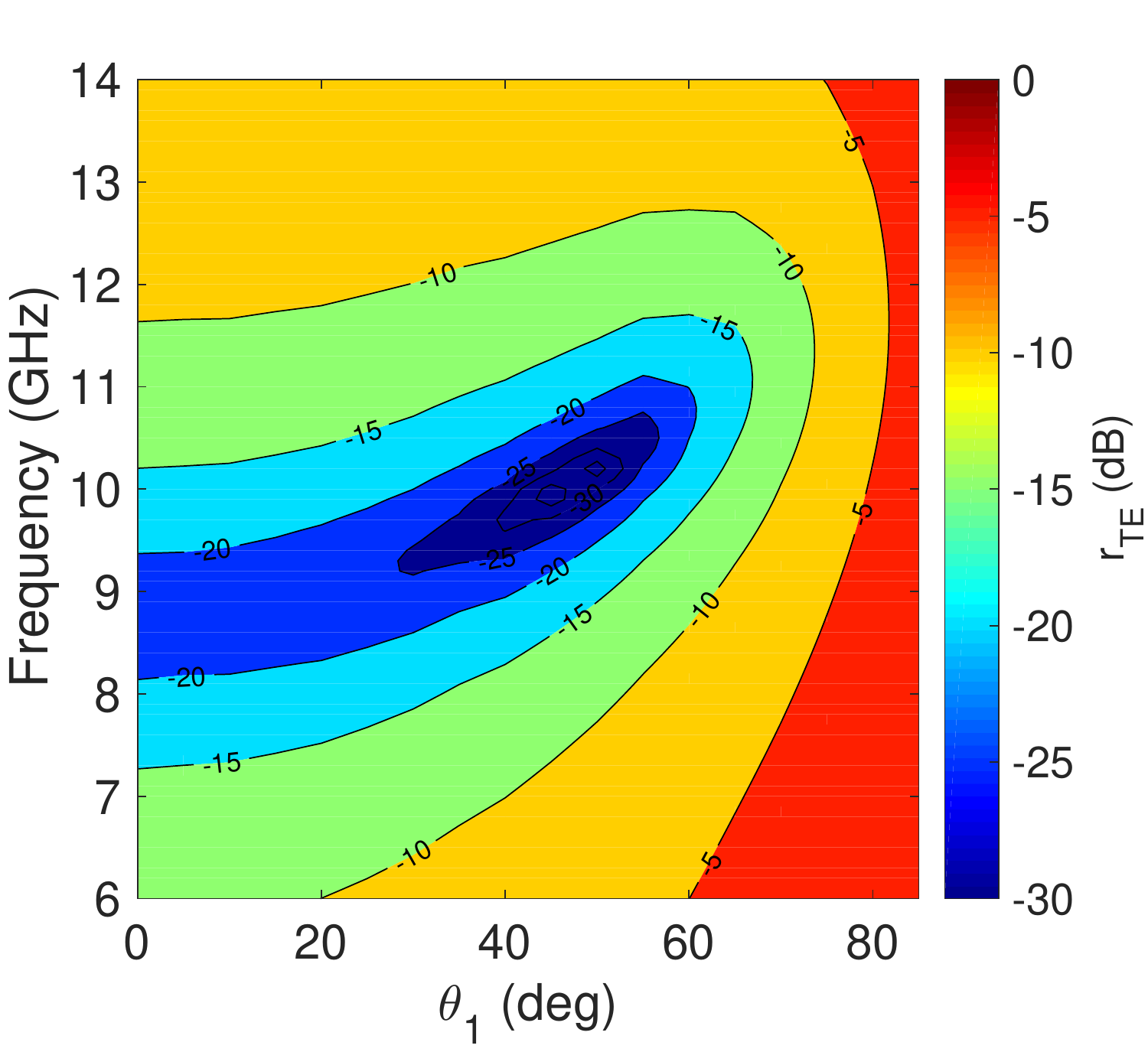}}
    \label{r_TE_theta_freq}
  \subfloat[]{%
        \includegraphics[width=0.3\linewidth]{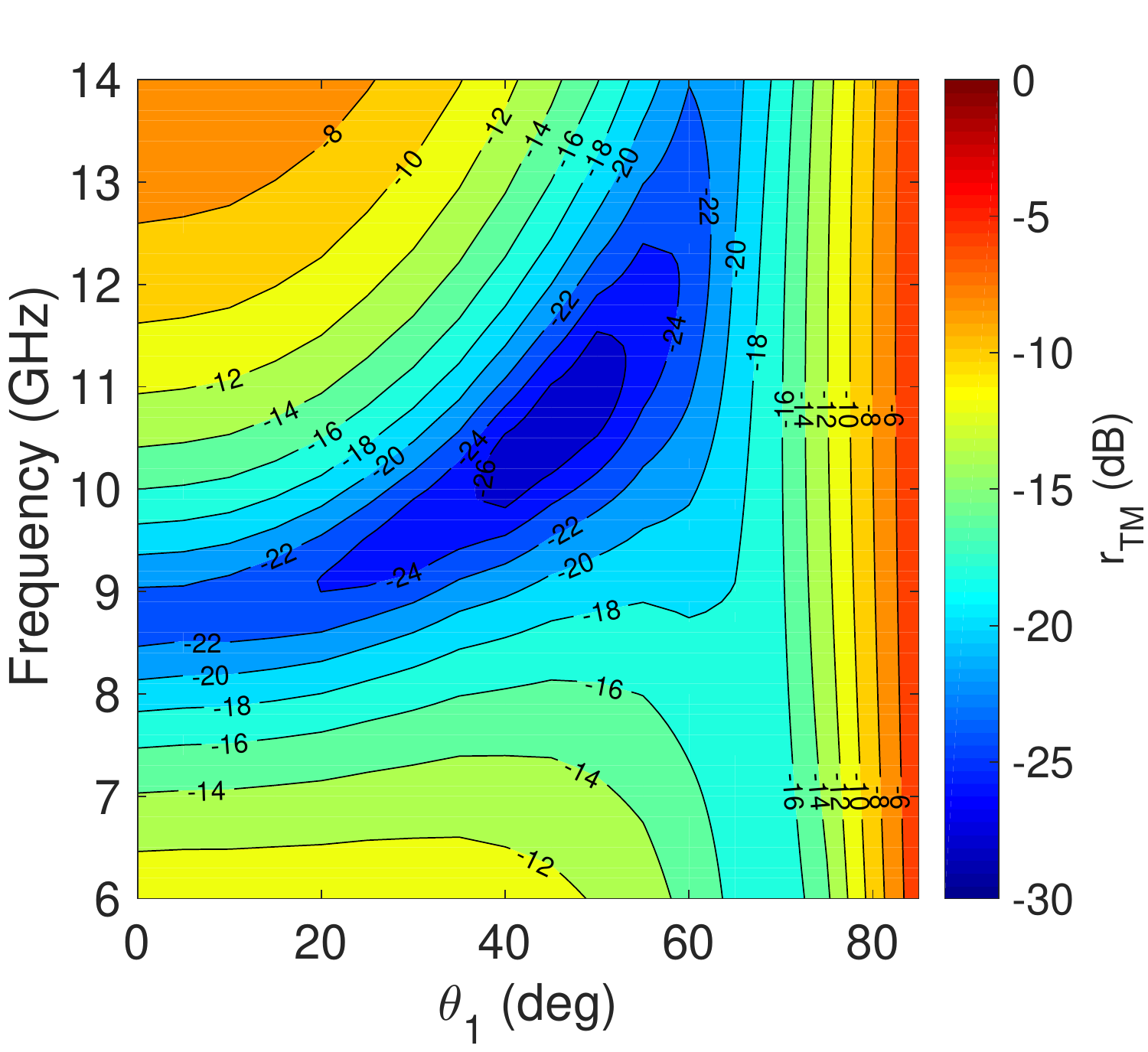}}
    \label{r_TM_theta_freq}
  \caption{{ Reflections are plotted against $\theta_1$ and frequency.} (a) TE polarization, (b) TM polarization. Due to the non-resonant nature of the MEUML, the matching performance is very wideband. At the design angle, the fractional bandwidths for a -20 dB reflection level are 15\% and 29\% for the TE and TM polarizations, respectively. }
  \label{fig:r_TE_TM_vs_theta_freq} 
\end{figure}

\begin{figure}[htb!]
    \centering
  \subfloat[]{%
       \includegraphics[width=0.3\linewidth]{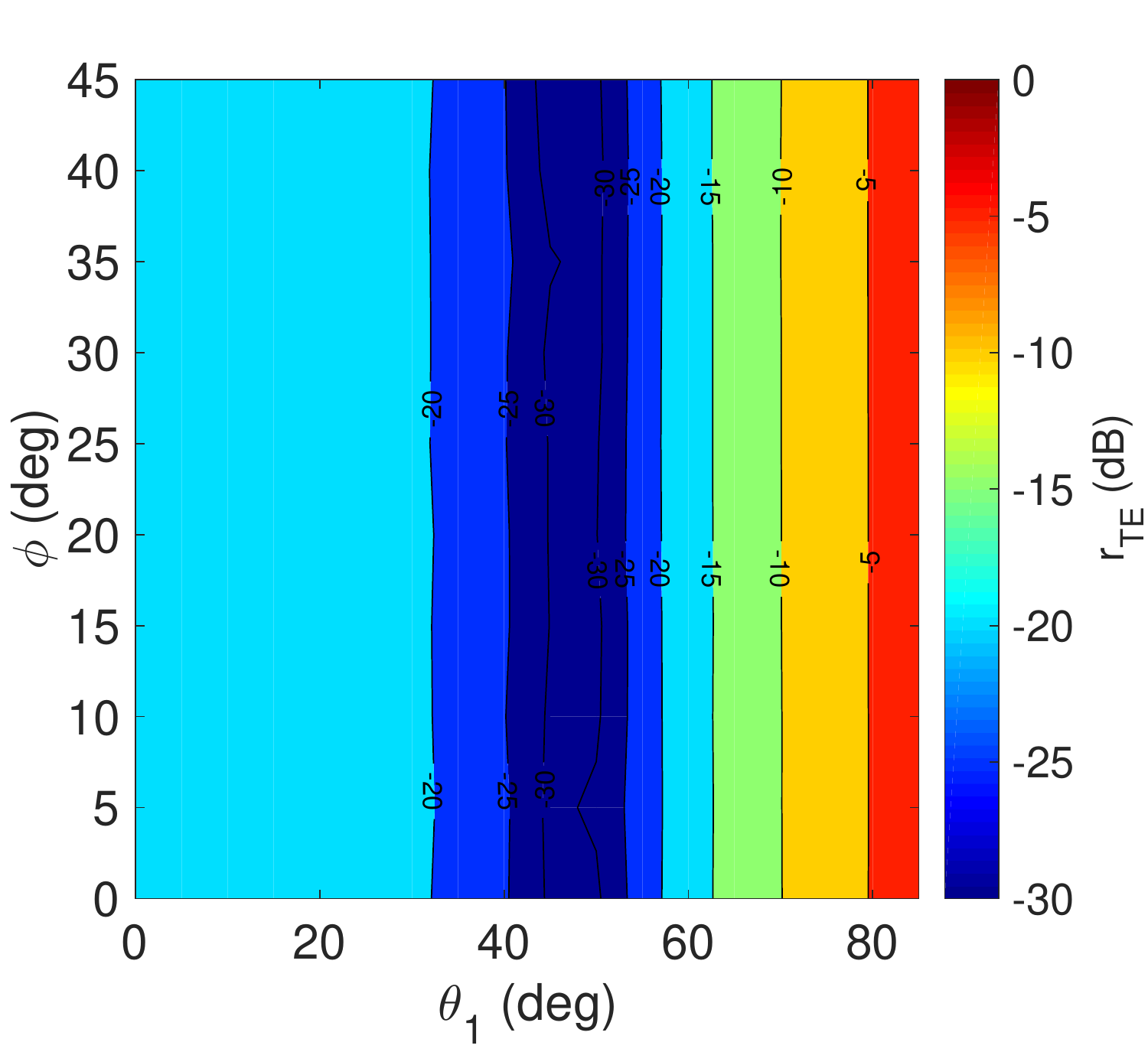}}
    \label{r_TE_theta_phi}
  \subfloat[]{%
        \includegraphics[width=0.3\linewidth]{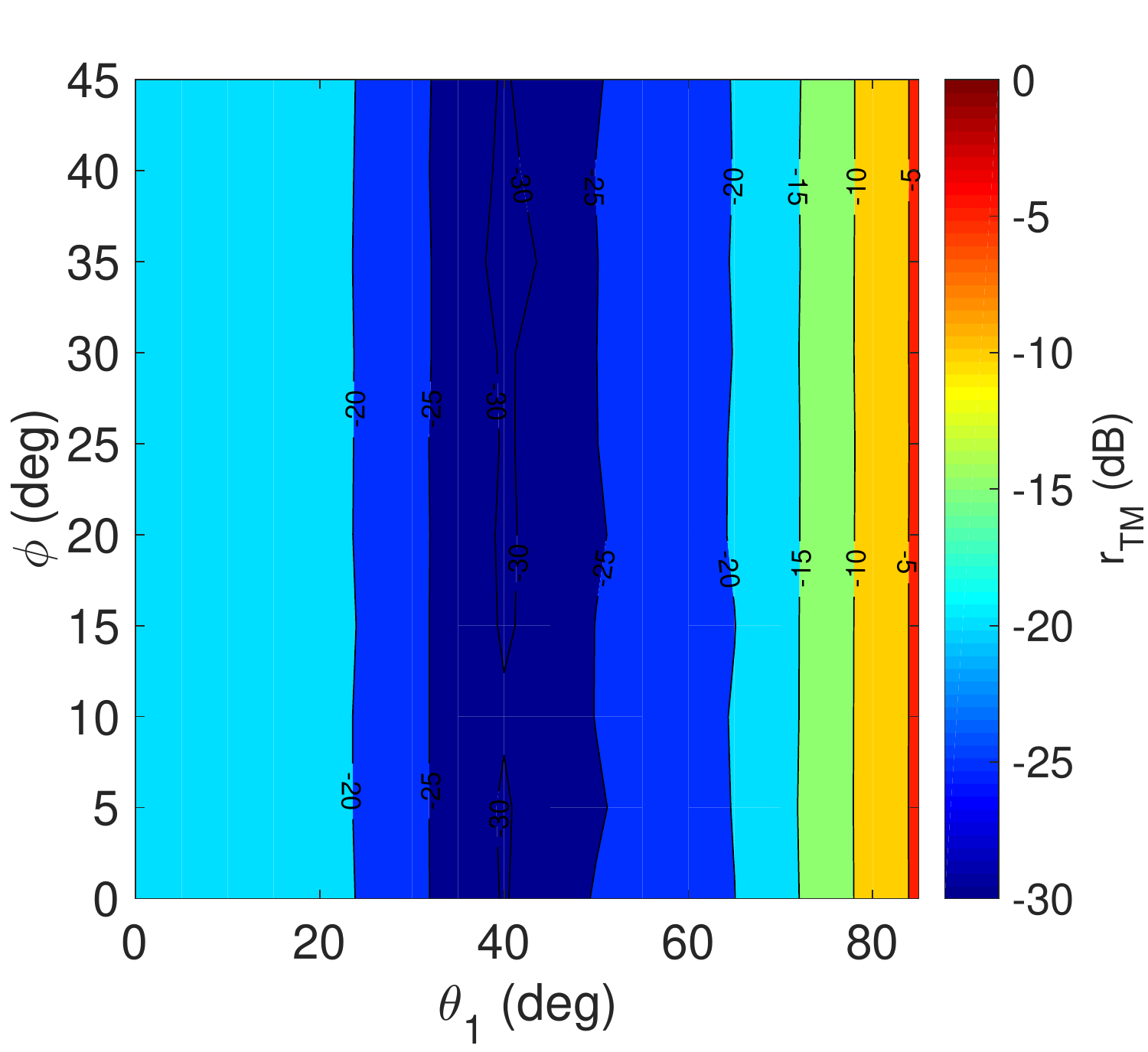}}
    \label{r_TM_theta_phi}
  \caption{{The azimuthal response of the MEUML.} (a) $r_{TE}$ and (b) $r_{TM}$ are plotted against $\theta_1$ and $\phi$, where $\phi$ is the azimuthal angle in the $x-y$ plane. Due to the symmetry of the unit cell, $\phi$ is only plotted from $0^\circ-45^\circ$. It is clear that the matching performance is independent of $\phi$. }
  \label{fig:r_TE_TM_vs_theta_phi} 
\end{figure}

The reflections for TE and TM-polarized waves are plotted against $\theta_1$ and frequency in Fig.~\ref{fig:r_TE_TM_vs_theta_freq}. Since the synthesized MEUML is not resonant, the matching performance is wideband. At the design angle, the fractional bandwidths for a -20 dB reflection reflection level are 15\% and 29\% for TE and TM polarization, respectively. From Fig.~\ref{fig:r_TE_TM_vs_theta_phi}, it is also clear that the matching performance is independent from $\phi$, which is the azimuthal angle in the $x-y$ plane. This is a result of the small unit cell size ($\lambda_0/6$) and the symmetrical geometries (circular rings and holes). The MEUML behaves as a well-homogenized slab with little material parameter variations in the tangential directions. The independence from $\phi$ is important for practical applications since the incoming signal may impinge at an arbitrary azimuthal angle. Matching variations in $\phi$ may lead to performance degradation.

\subsection{MEUML based radome}\label{sec:radome_design}
As was mentioned previously, for practical microwave applications, the matching layer is rarely used to match from air to a semi-infinite thick substrate. Rather, the matching is from air to air. In such a scenario, the matching layer is essentially a radome. {\color{black} Radomes are mechanically robust and act as RF (radio frequency) transparent enclosures that protect expensive radar or telecommunication apparatus from environmental effects. They have been widely used in weather and airborne radars. With emerging 5G-telecommunications \cite{boccardi2014five} and autonomous driving \cite{6127923}, phased arrays that operate at millimeter-wave frequencies have been widely adopted. Since the free-space loss at such high frequencies can be significant, a proper radome design is critical for minimizing the additional loss due to reflections. For many practical considerations, the radome should be low-profile such that the overall system is compact \cite{Automotive_radome_rib}. This inevitably requires the radome to maintain a good transmission as the phased array scans over a wide angular range.} To achieve a good transmission over a wide angular range for both polarizations, a conventional radome is usually a sandwich type of construction \cite{cady1948radar}, which consists of two high/low-index thin dielectric sheets separated by a low/high index dielectric spacer. Minimization of the total reflection can be achieved by the mutual cancellation of the reflections between the sheets. The indices of refraction and the thicknesses of the sheets, and the spacer, are extensively optimized to achieve the best angular performance for both polarizations. However, since the materials used are generally isotropic and non-magnetic, it is difficult to achieve the same performance for both polarizations. The performance for TM-polarizations is usually better due to the existence of a Brewster's angle. 

With the proposed MEUML concept, one gains control in the longitudinal and tangential permittivities and permeabilities, which enables additional degrees of freedom in the radome design. To demonstrate this concept, a radome is constructed by having a thin RO3010 substrate sandwiched between two MEUMLs as previously described.
The total reflection of the sandwich radome is calculated by the transfer matrix method (See Appendix~\ref{sec:TMM analysis of the MEUML radome}) and it is given by \eqref{eq:sandwich_reflection}
\begin{equation}
r_{sandwich}=\frac{r_{02}e^{i\phi_2}-r_{02}e^{-i\phi_2}(t_{20}t_{02}-r_{02}r_{20})}{e^{i\phi_2}-e^{-i\phi_2}r_{02}r_{20}}
\label{eq:sandwich_reflection}
\end{equation}
where $t_{02}$ is the total transmission from the semi-infinite air region to the semi-infinite substrate; $t_{20}$ is the total transmission from the semi-infinite substrate to the semi-infinite air region; $r_{02}$/$r_{20}$ is the total reflection with the incidence from the semi-infinite air/substrate. Thus, $r_{02}$ and $t_{02}$ are equivalent to $r$ and $t$ in Fig.~\ref{fig:TandRandloss}(a) and Fig.~\ref{fig:TandRandloss}(b), respectively.

\begin{figure}[htb!]
\centering
\includegraphics[width=0.5\columnwidth]{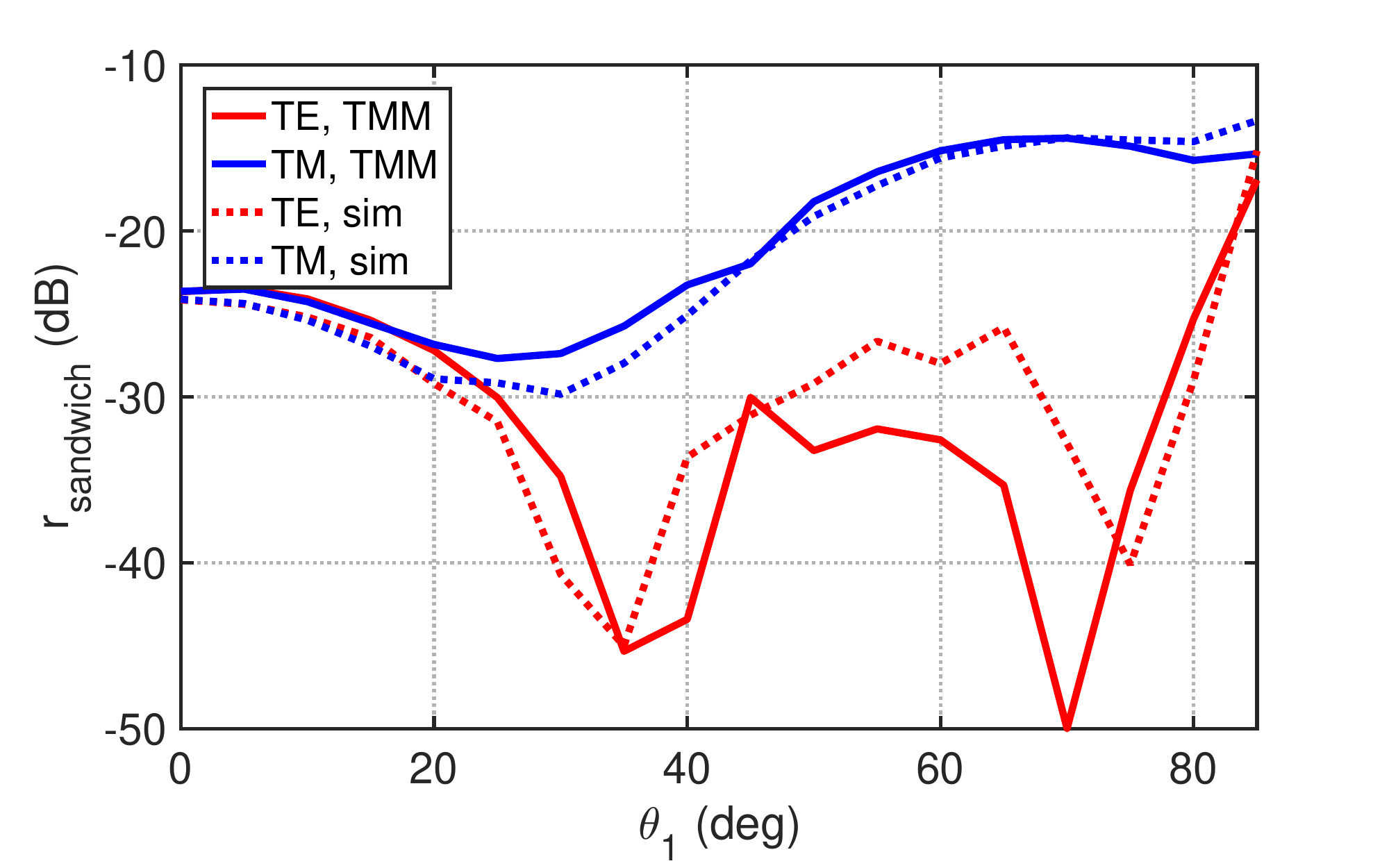}
\caption{{The total reflection of the MEUML radome is plotted against $\theta_1$.} Excellent agreement is observed between the TMM calculation and the simulation results. The reflections for both polarizations stay below -14 dB for the entire angular range of $0^\circ-85^\circ$.}
\label{fig:stack_d3010_1p27mm_reflection}
\end{figure}

\begin{figure}[htb!] 
    \centering
		%
  \subfloat[]{%
         \includegraphics[width=0.3\linewidth]{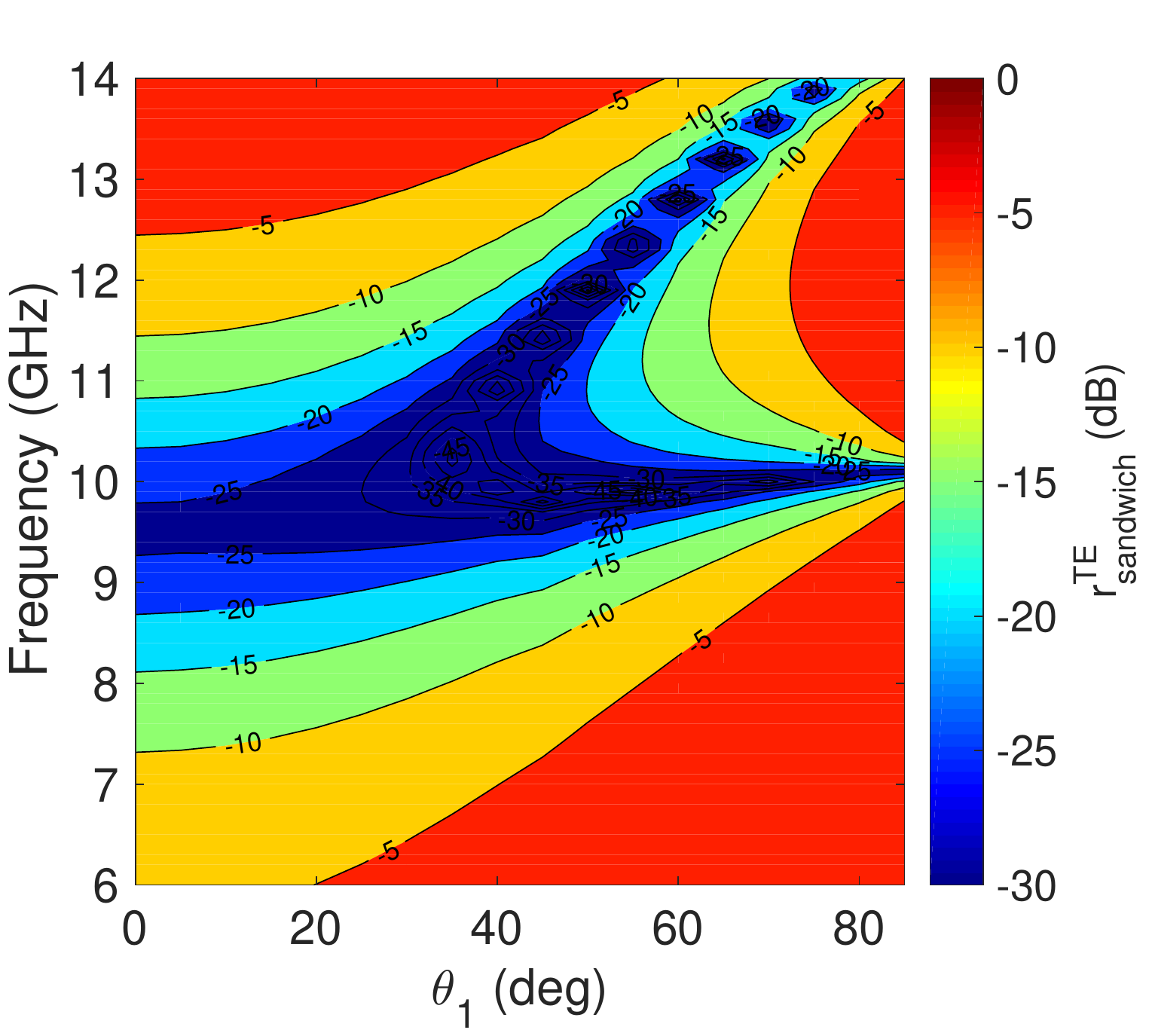}}
    \label{fig:stack_r_TE_vs_theta_freq}
  \subfloat[]{%
        \includegraphics[width=0.3\linewidth]{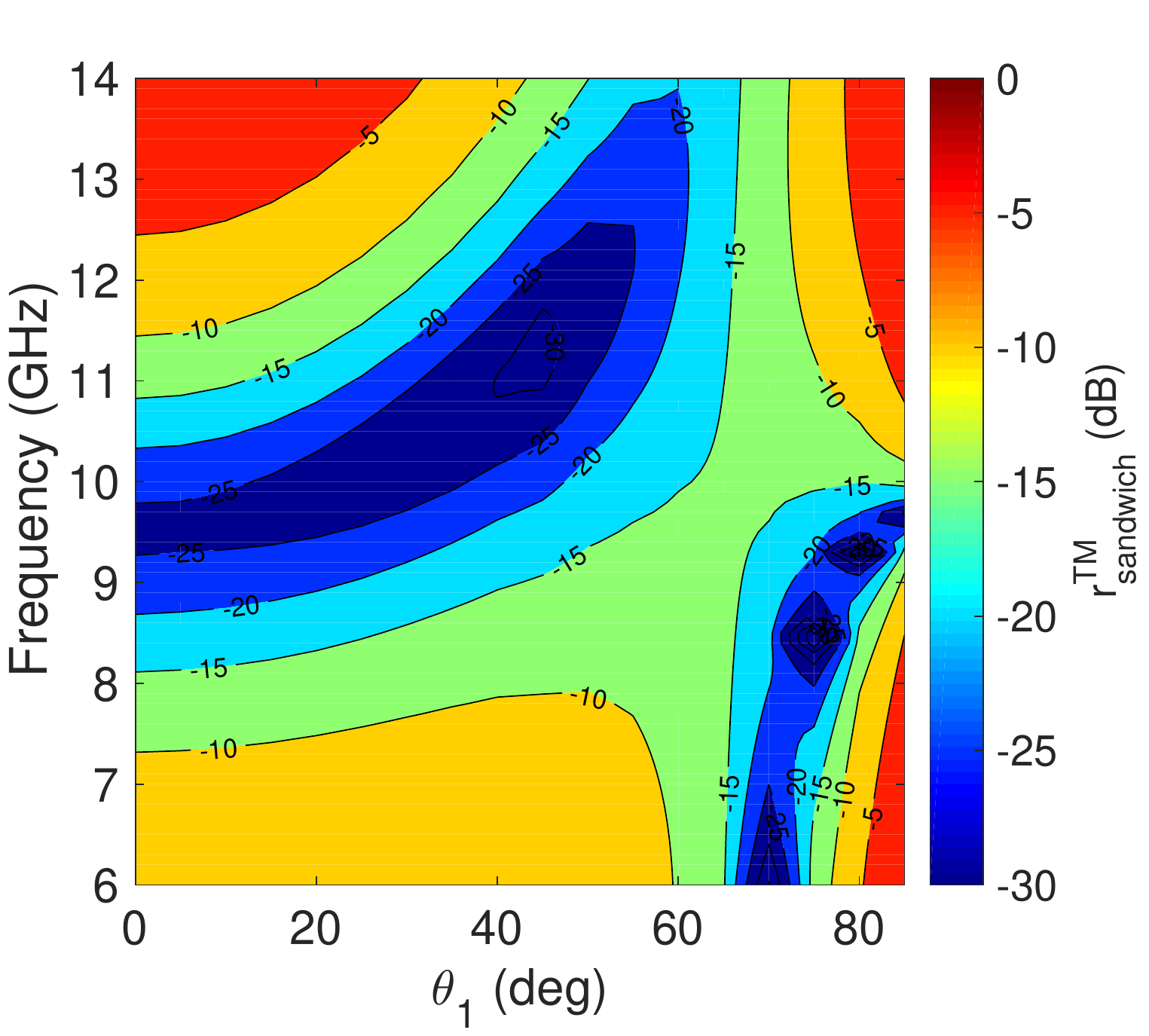}}
    \label{fig:stack_r_TM_vs_theta_freq}
		
  \caption{{Reflection levels of the MEUML radome.}  The simulated total reflections of the MEUML radome are plotted against $\theta_1$ and frequency for (a) TE polarization and (b) TM polarization. The matching performance is wideband as expected. The bandwidths for maintaining less than -10 dB reflection for the entire range of $0^\circ-85^\circ$ are 0.3 GHz and 1.1 GHz for the TE and TM polarizations, respectively. }
  \label{fig:MEUML_reflection} 
\end{figure}

From \eqref{eq:sandwich_reflection}, we immediately notice that if the MEUML achieves perfect matching between the semi-infinite air and the semi-infinite substrate ($r_{02}=0$), the total reflection of the sandwich structure is immediately zero and it is independent of the thickness (which is related to $\phi_2$) of the RO3010 substrate. 
In addition, as long as $r_{02}$ is small, $r_{sandwich}$ is also small and has a weak dependance on the substrate thickness. Thus, from Fig.~\ref{fig:TandRandloss}(a), we can infer that the sandwich structure will have a small reflection between $0^\circ-60^\circ$ as $r_{02}$ is very small in this angular range. To reduce the reflection of the sandwich structure beyond $60^\circ$, the thickness of the RO3010 can be optimized. Due to the weak dependance on the substrate thickness for angles between $0^\circ-60^\circ$, the optimization process will not introduce too much performance degradation in this angular range. 
With an optimal thickness of 1.25mm (See Appendix~\ref{sec:TMM analysis of the MEUML radome}), the maximum reflections for both TE and TM polarizations are -14.6 dB, which correspond to a reflected power of 3.5\%. For fabrication purposes, the RO3010 substrate is chosen with a standard thickness of 1.27mm. The resulting radome is very compact and has an overall thickness of 10.8mm, which is about $\lambda_0/3$. Fig.~\ref{fig:stack_d3010_1p27mm_reflection} shows the total reflection of the MEUML radome as a function of the incident angle. Very good agreement is observed between the calculated and simulated results. To the author's best knowledge, the performance of maintaining less than 5\% reflected power over $0^\circ-85^\circ$ for both TE and TM polarizations is one of the best results reported so far in the literature. In addition, the radome is wideband as shown in Fig.~\ref{fig:MEUML_reflection}(a) and Fig.~\ref{fig:MEUML_reflection}(b). For TE-polarized waves, the bandwidth is $9.95-10.25$ GHz such that the reflection stays below -10 dB for the entire angular range of $0^\circ-85^\circ$. For TM-polarized waves, the -10~dB bandwidth is $9.1-10.2$ GHz. For applications that require a smaller angular range, e.g., $0^\circ-60^\circ$, the -10 dB reflection bandwidths become 2.5 GHz and 3.6 GHz for TE and TM polarizations, respectively.

\begin{figure}[htb!] 
  \centering
  \subfloat[]{%
         \includegraphics[width=0.3\linewidth]{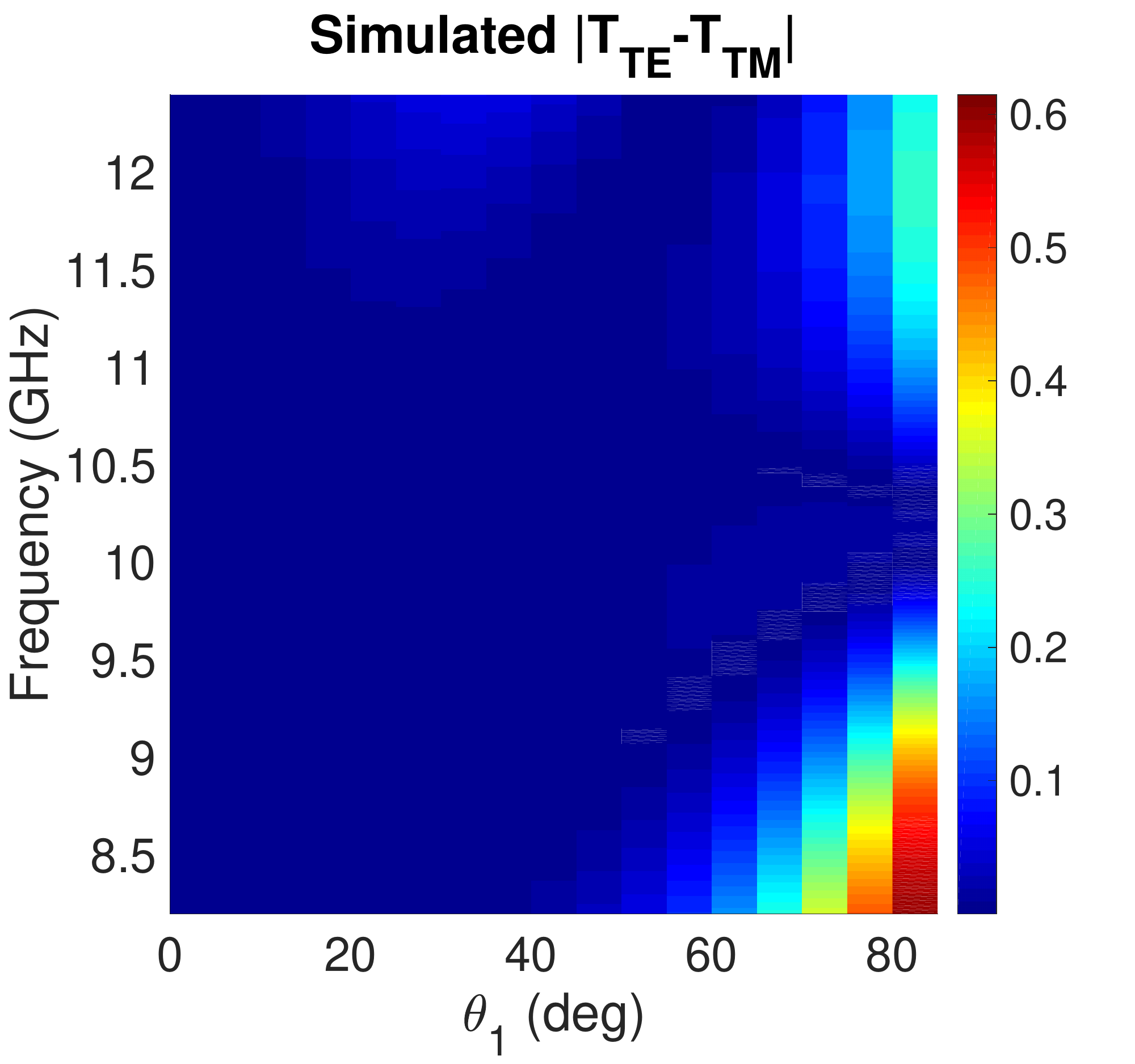}}
    \label{fig:Simulated_UMML_radome_transmission_mag_difference}
  \subfloat[]{%
        \includegraphics[width=0.3\linewidth]{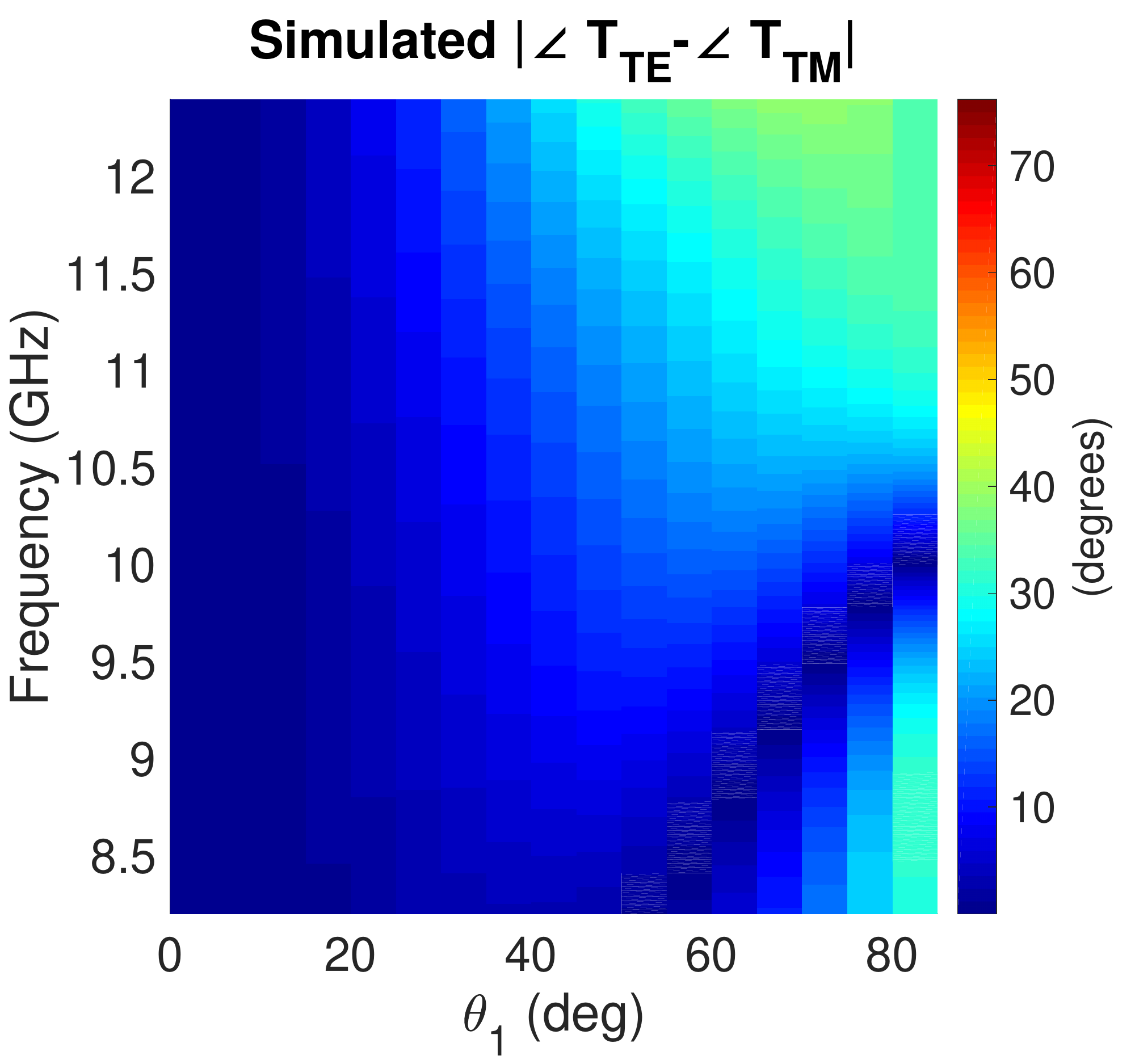}}
    \label{fig:Simulated_UMML_radome_transmission_phase_difference}
		
  \caption{{\color{black}The simulated transmission (a) magnitude difference and (b) phase difference between the two polarizations are plotted. Maintaining small magnitude and phase difference is important for the polarization fidelity. Near the design frequency, the maximum magnitude difference is less than 5\% and the maximum phase difference is less than $16^\circ$.}}
  \label{fig:MEUML_transmission difference} 
\end{figure}

{\color{black}
For optimal radome performance, it is desirable to minimize the magnitude and phase differences to maintain a good polarization fidelity. If the incoming linearly polarized signal becomes elliptically polarized after passing the radome, the gain of the linearly polarized receiver would drop. Furthermore, for sources placed behind the radome, it is also desirable that the magnitude and phase has small variations vs. the scan angle. Otherwise, the radiation pattern would be distorted. Fig.~\ref{fig:MEUML_transmission difference} shows the magnitude and phase differences between the two transmitted polarizations. We can see that near the design frequency, the maximum magnitude difference is less than 5\% and the maximum phase difference is less than $16^\circ$ for angles between $0^\circ-85^\circ$.
}

\section{Experiments}\label{sec:experiment}
Due to the limitation in fabrication and material availability, it is not possible to characterize a MEUML sandwiched by a semi-infinite air region and a semi-infinite (or a very thick) substrate. Thus,  we choose to only characterize the radome which in fact is a more relevant structure from the applications perspective.

To characterize the radome, we performed a wideband far-field measurement with the setup shown in Fig.~\ref{fig:FarField_measurement} (characterization details are described in the Appendix~\ref{sec:MEUML Fabrication and Measurement}).  Fig.~\ref{fig:FarField_measurement}(d)-(g) compare the simulated and measured radome transmission. The simulated and measured results are in good agreement. The radome is wideband; it maintains near unity transmission between $8.2-12.4$~GHz for angles up to $80^\circ$. However, the accurate transmission of the radome cannot be obtained for the following reasons: First, the far-field measurement setup has an accuracy no better than 0.5 dB. Second, the radome was placed in the near field of the small receiving antenna, which may distort the radiation pattern of the antenna. Thus, when calculating the transmission of the radome from the measured gain difference, the transmission can be inaccurate and sometimes exceed unity. Third, the radome has a finite size; thus, diffraction occurring at the edges of the MEUML may introduce additional inaccuracies, especially at extreme angles. 

To obtain a more accurate radome transmission, we conducted a quasi-optical measurement as shown in Fig.~\ref{fig:QO_measurement} (characterization details are described in the Appendix~\ref{sec:MEUML Fabrication and Measurement}).  Since the quasi-optical setup is narrowband, the radome was measured only at 10~GHz. The simulated and measured transmissions are shown in Fig.~\ref{fig:QO_measurement}(c). A very good agreement between the simulated and measured results can be observed for angles between $0^\circ-65^\circ$. The measured results drop more sharply beyond $65^\circ$. This is due to the spill-over loss. As illustrated in Fig.~\ref{fig:QO_measurement} (a), at $70^\circ$,  the projection of the finite-size (30cm long) sandwich structure onto the beam-waist plane is smaller than the Gaussian beam-waist. Thus, near $70^\circ$, diffraction occurring at the edges of the radome leads to spill-over loss and results in a decrease in the measured transmission. Nevertheless, the good agreement between the simulated and measured results for less oblique angles validates the theory and the design of the MEUML based radome.

\begin{figure*}[htb!]
    \centering
				\subfloat[]{%
				 \includegraphics[width=0.3\linewidth]{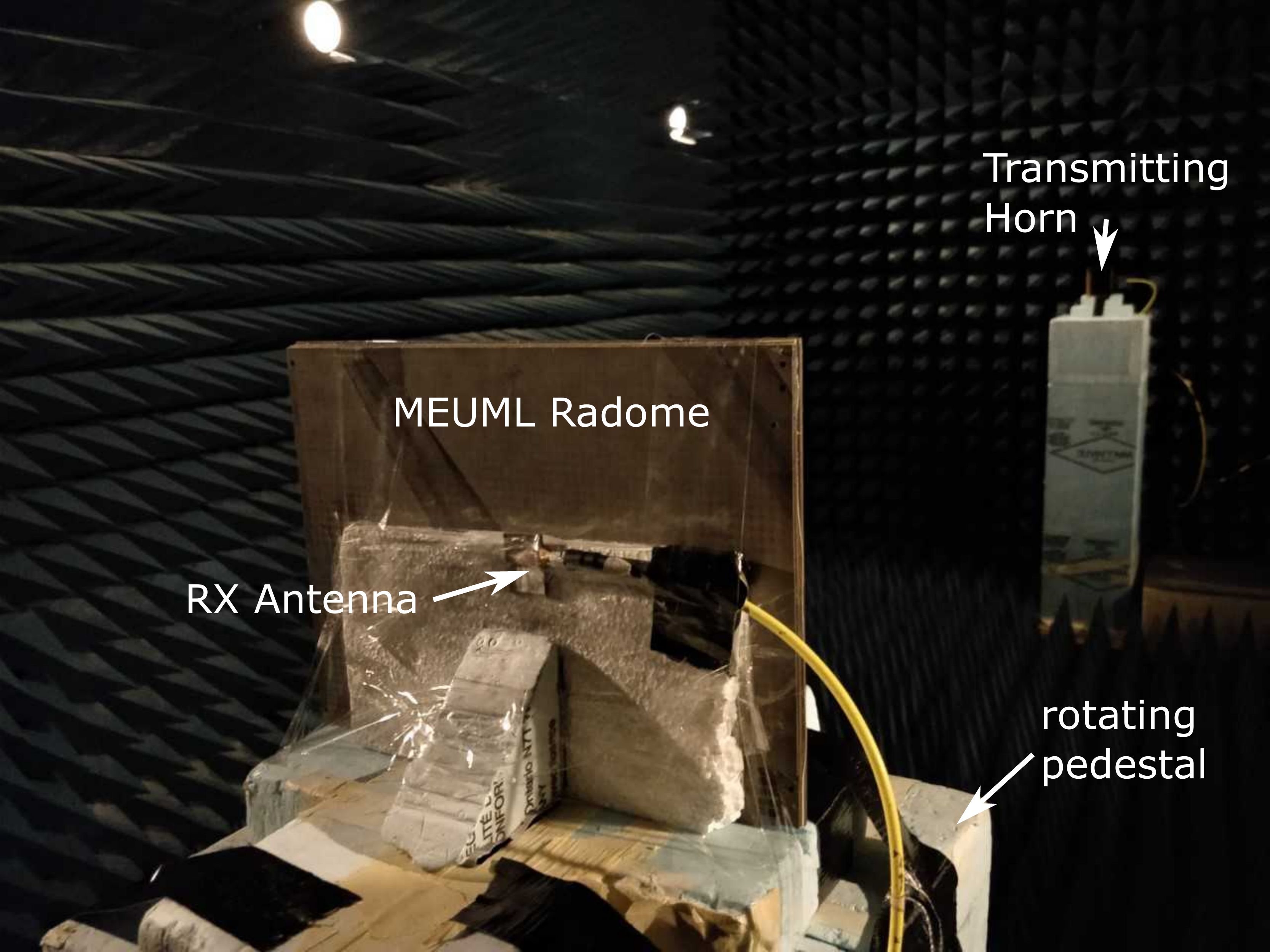}}
			\label{fig:FF_measurement}
					\subfloat[]{%
				 \includegraphics[width=0.35\linewidth]{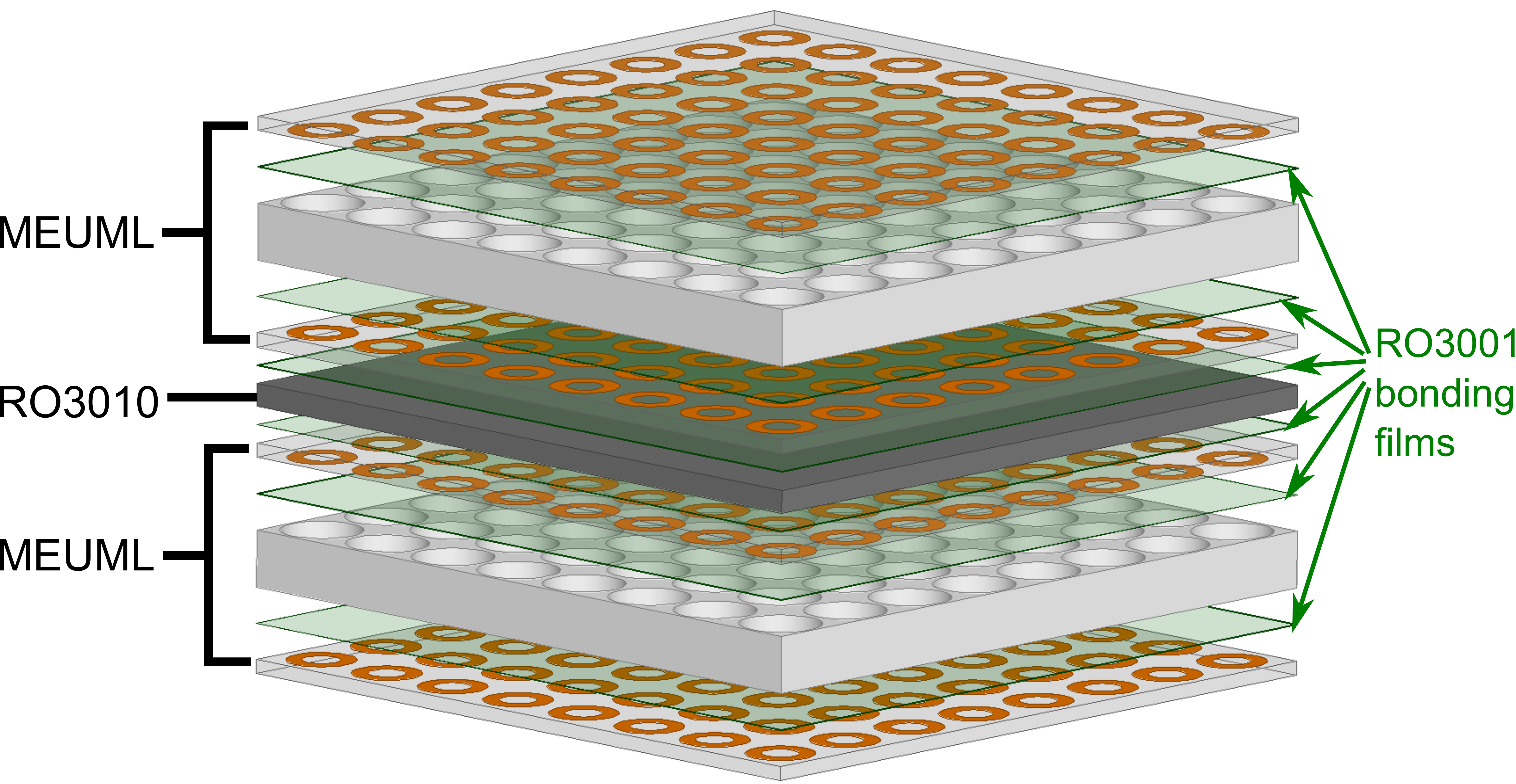}}
			\label{fig:MEUML_radome_exploded_view}
					\subfloat[]{%
				 \includegraphics[width=0.2\linewidth]{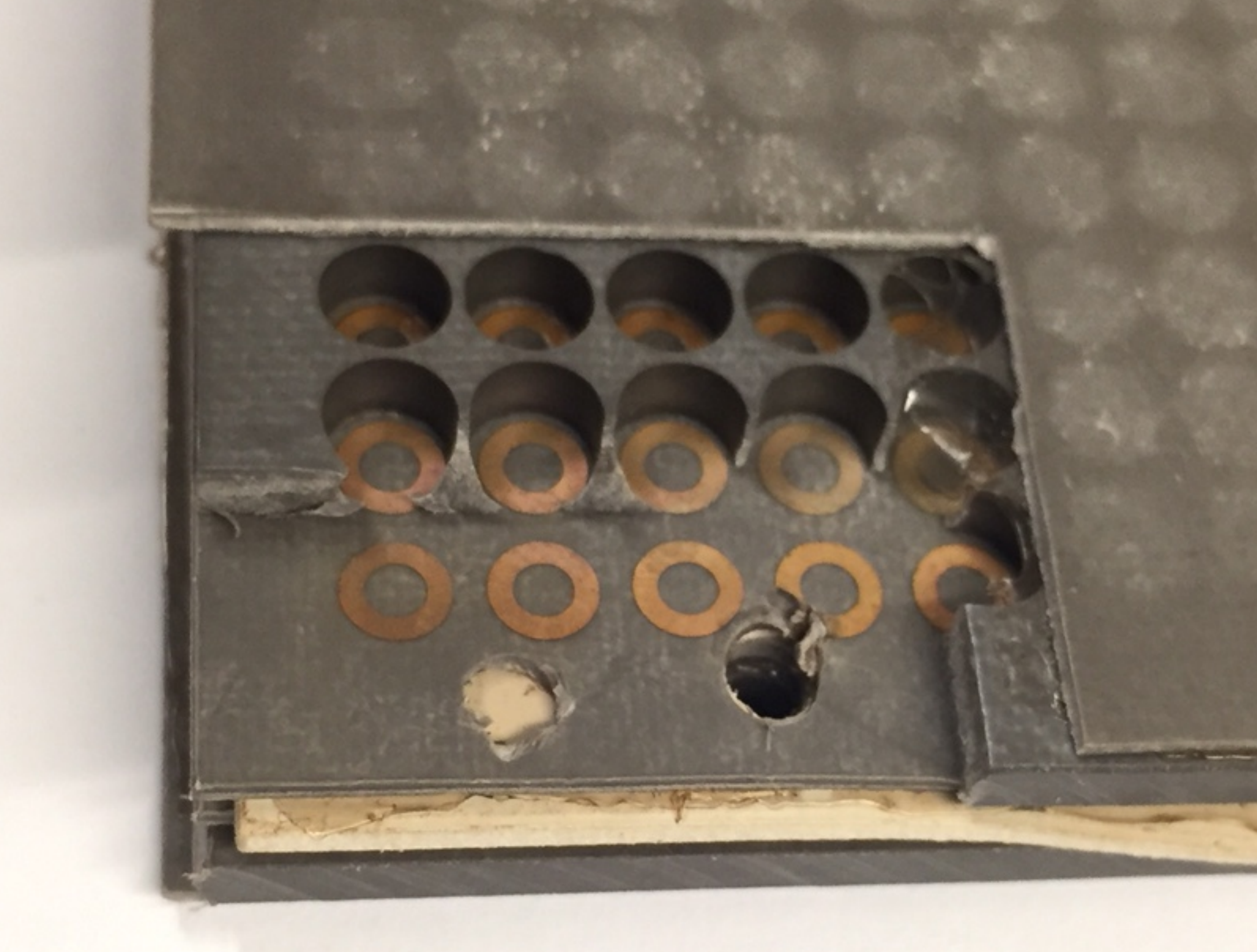}}
			\label{fig:MEUML_radome_fabricate_feature}

		\subfloat[]{%
				 \includegraphics[width=0.245\linewidth]{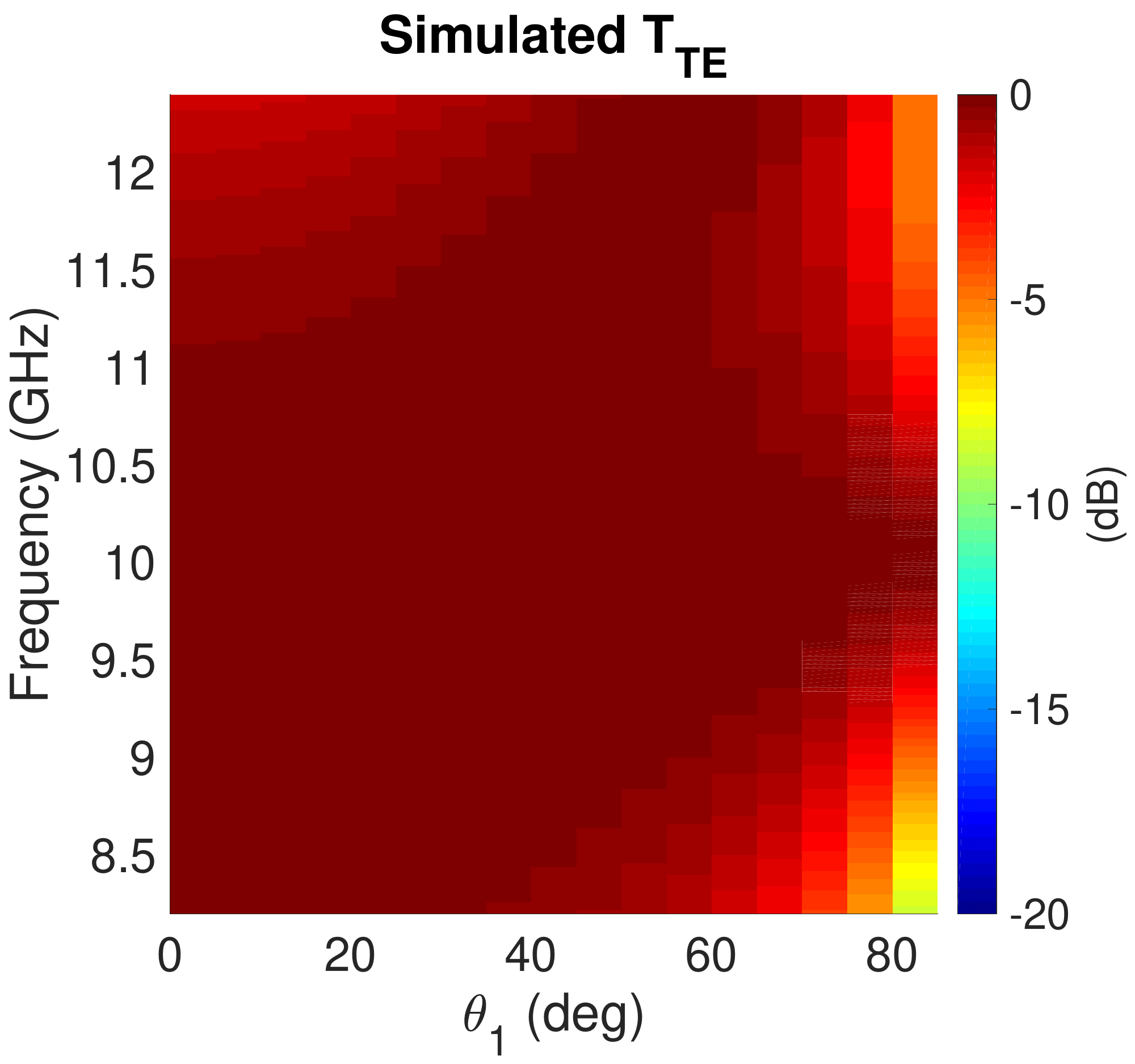}}
			\label{fig:Simulated_MEUML_tTE_dB}
		\subfloat[]{%
					\includegraphics[width=0.245\linewidth]{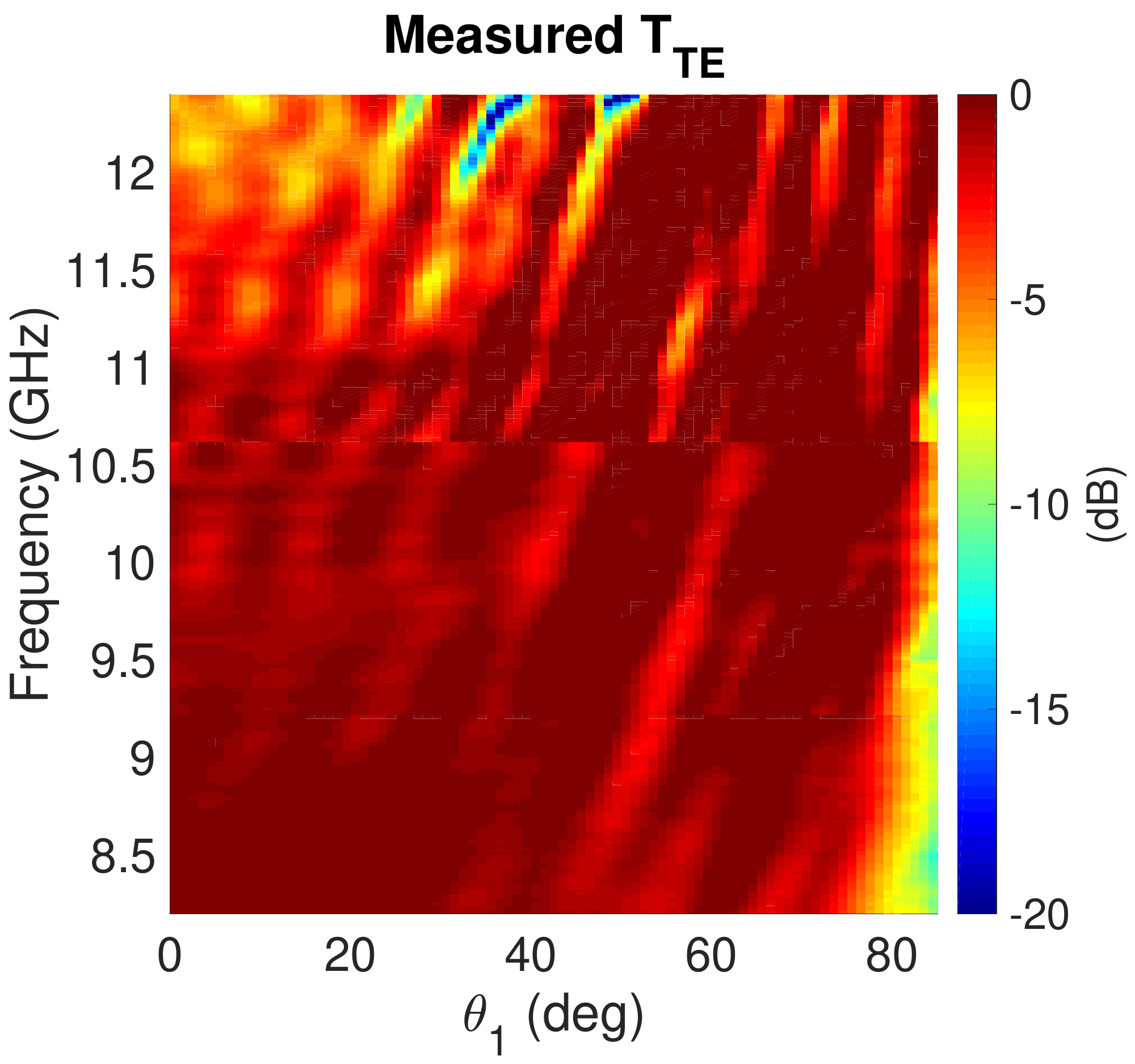}}
			\label{fig:Measured_MEUML_tTE_dB}
		  \subfloat[]{%
       \includegraphics[width=0.245\linewidth]{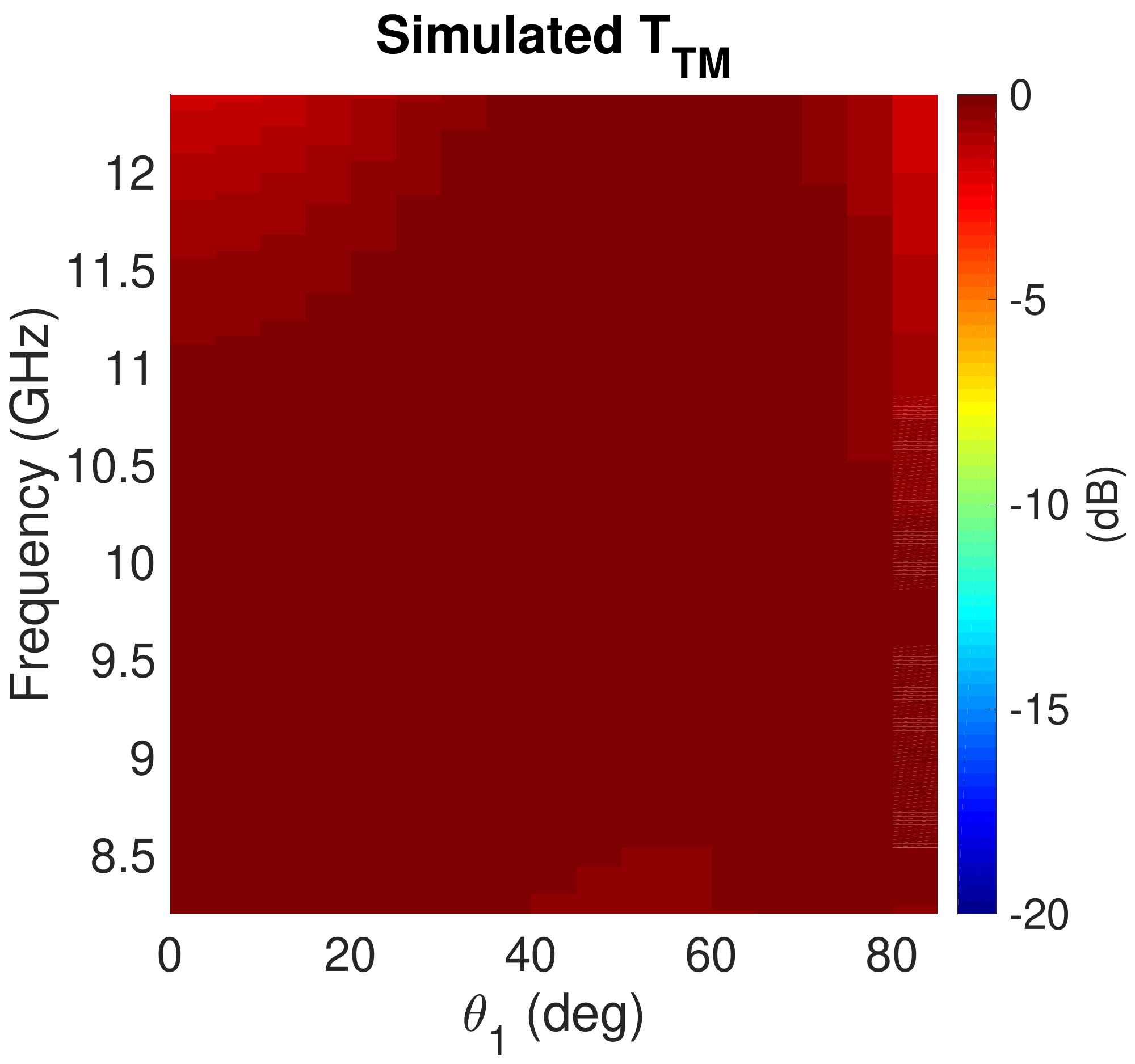}}
    \label{fig:Simulated_MEUML_tTM_dB}
  \subfloat[]{%
        \includegraphics[width=0.245\linewidth]{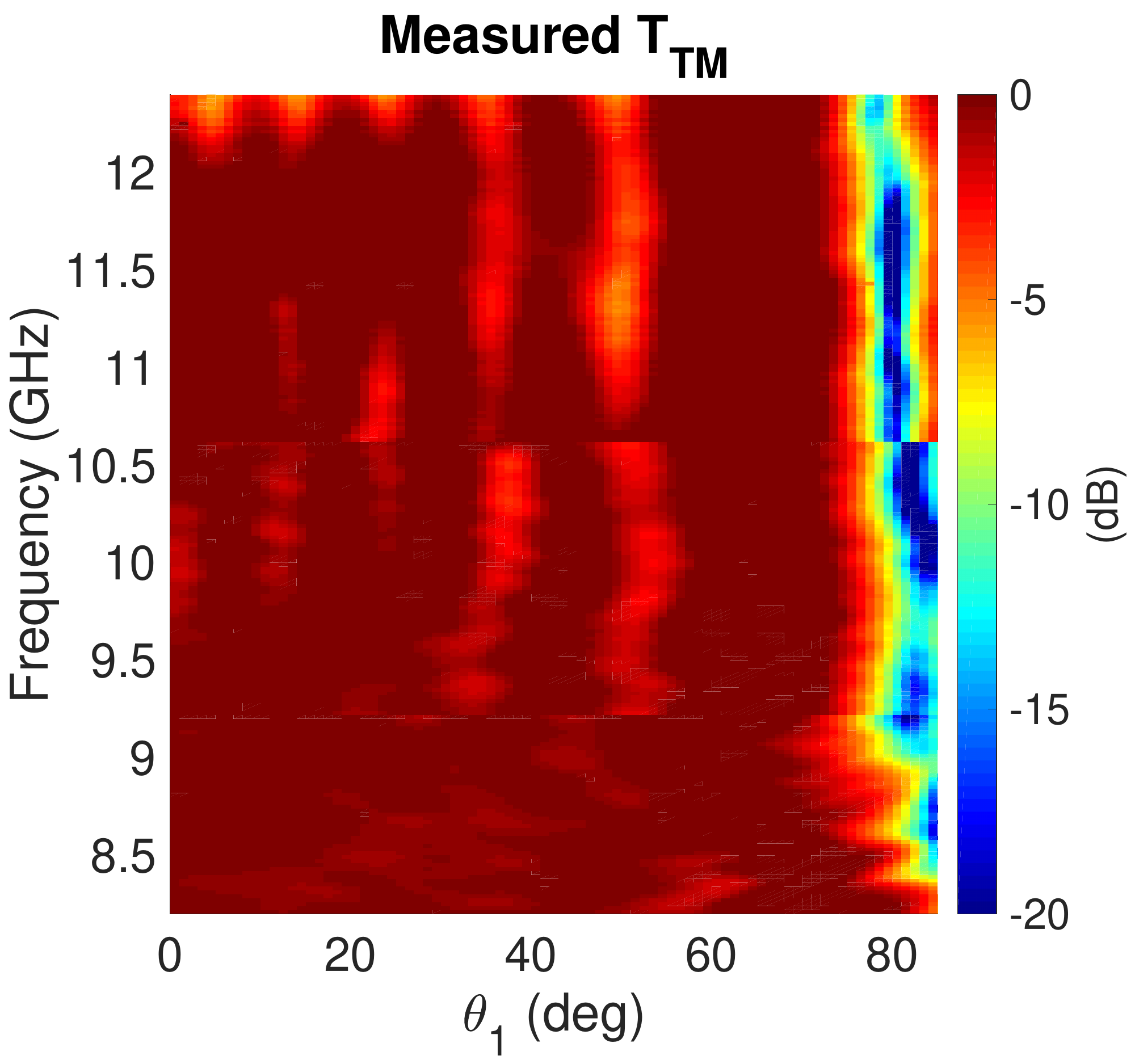}}
    \label{fig:Measured_MEUML_tTM_dB}
		
  \caption{{Far-field measurement of the MEUML radome. }(a) shows the far-field measurement setup. A receiving antenna is placed behind the MEUML radome with a 1.5~cm thick foam as a spacer. Two gain measurements are conducted with and without the radome in front of the antenna. The transmission of the radome can be extracted from the measured gain difference. As the pedestal rotates, the transmission of the radome can be measured against the incident angle. The transmitting pyramidal horn can be rotated such that the radiated electric field is polarized either horizontally or vertically. Thus, the transmission of the radome for TE or TM-polarized waves can be measured accordingly. (b) shows the expanded view of the MEUML radome. For details on fabrication, please see Appendix~\ref{sec:MEUML Fabrication and Measurement}. (c) shows the detailed features of the fabricated prototype. (d) and (e) compare the simulated and measured transmission for TE polarization. (f) and (g) compare the simulated and measured transmission for TM polarization. The simulated and measured results are in good agreement. The radome maintains near unity transmission for angles up to $80^\circ$ and for the entire $8.2-12.4$~GHz.}
  \label{fig:FarField_measurement} 
\end{figure*}

\begin{figure*}[htb!]
  \centering
  \subfloat[]{%
        \includegraphics[width=0.6\linewidth]{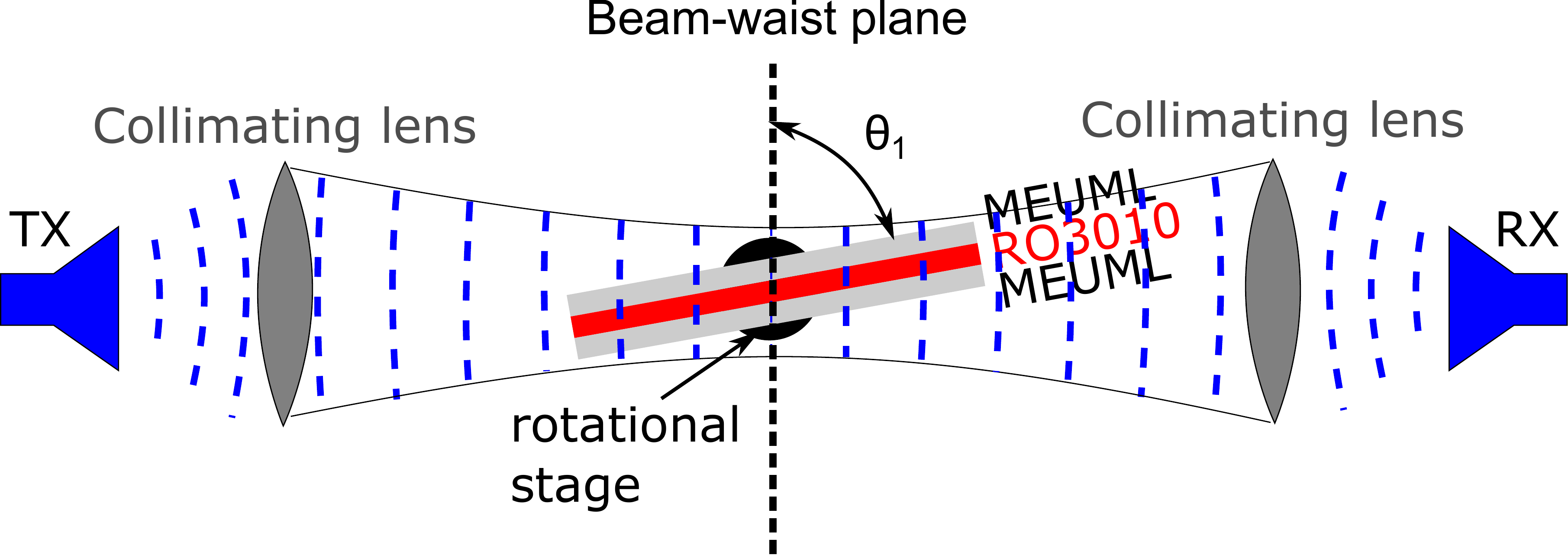}}
    \label{fig:experimental_setup_drawing}
		
		  \subfloat[]{%
        \includegraphics[width=0.43\linewidth]{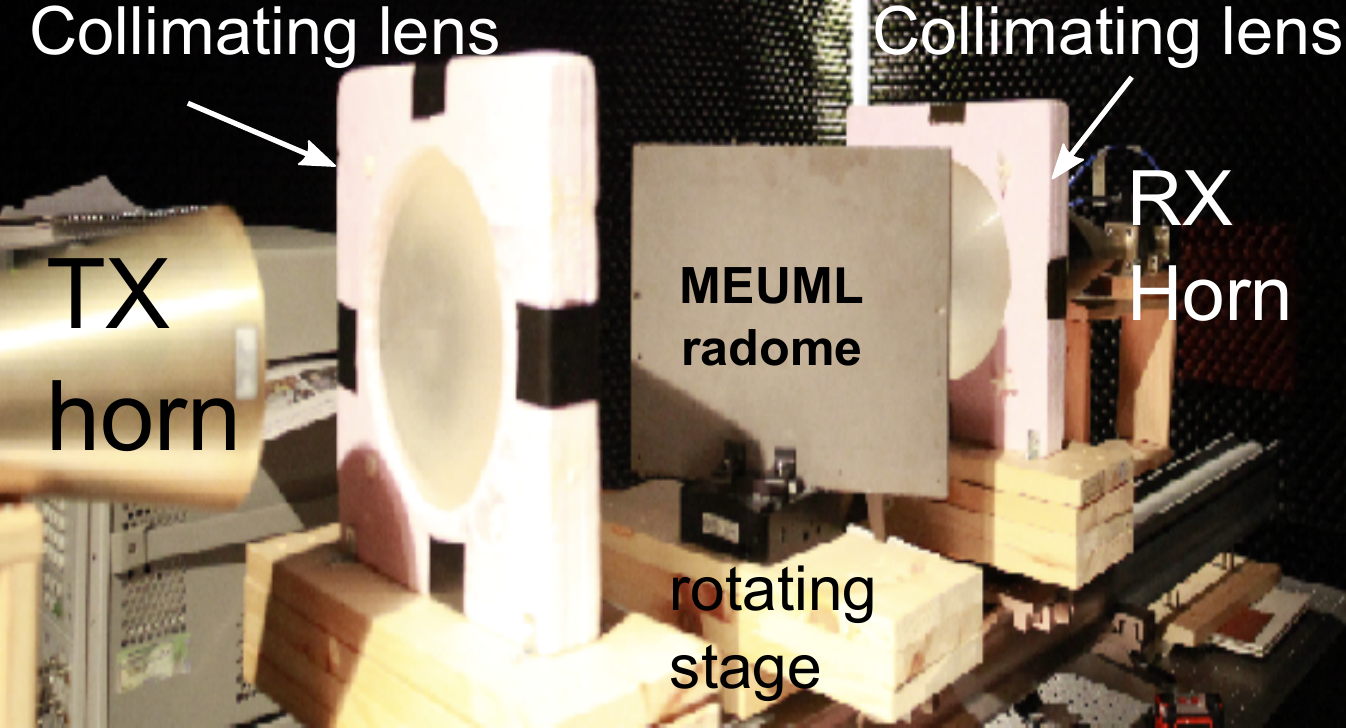}}
    \label{fig:Measurement_setup_shot}
		  \subfloat[]{%
        \includegraphics[width=0.4\linewidth]{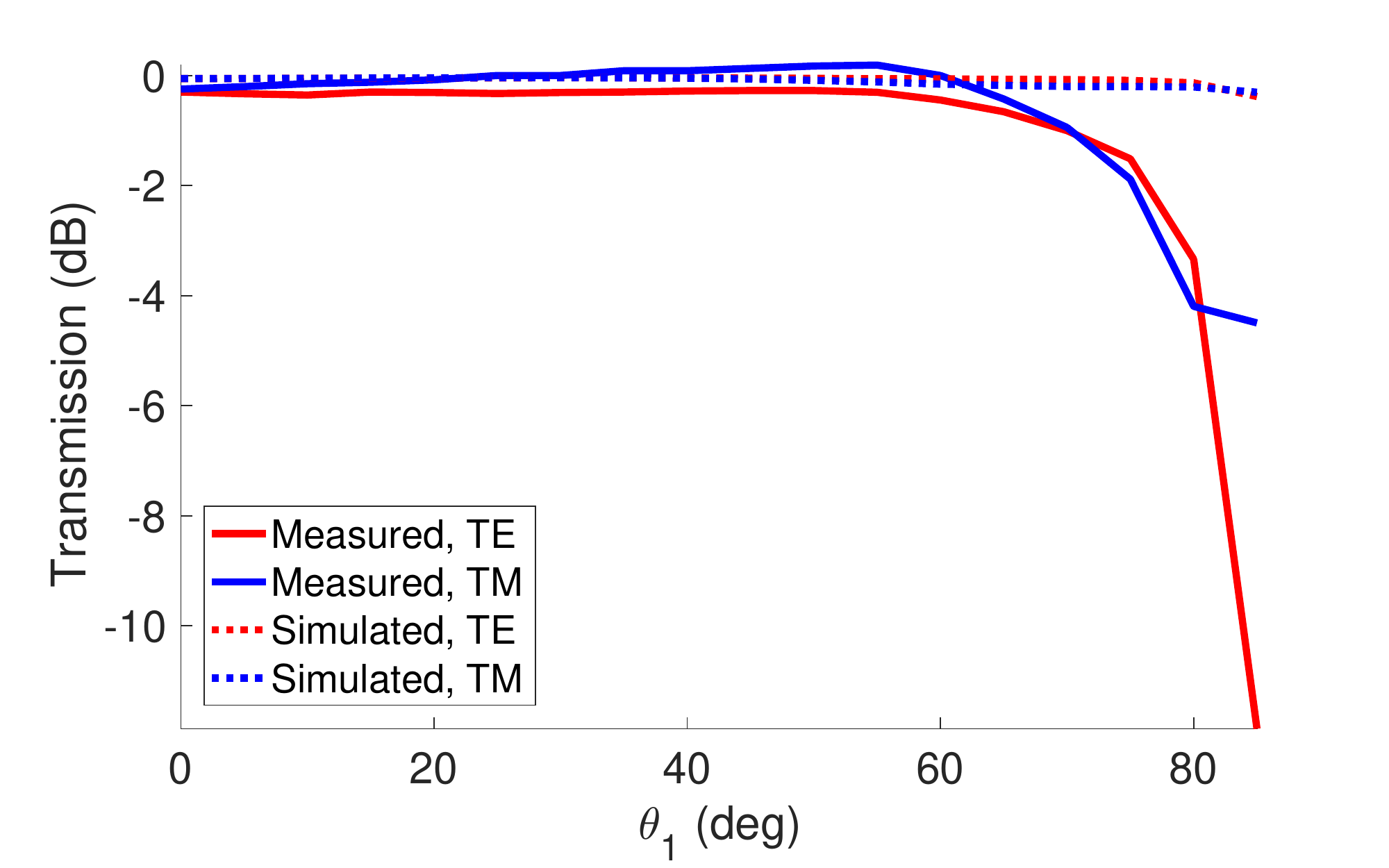}}
    \label{fig:stack_sim_vs_meas_result}

  \caption{{Quasi-optical measurement of the MEUML radome} (a) illustrates the quasi-optical characterization of the MEUML radome. Instead of measuring the total reflection, the total transmission is measured as the radome rotates. The radome is placed at the beam-waist plane such that the wavefront impinging on the radome is planar. (b) shows the quasi-optical setup. Since this is a narrow-band system, the measurement is performed only at 10~GHz.  (c) compares simulated and measured transmission of the MEUML radome for both TE and TM polarizations at 10~GHz.  Good agreements between the results can be observed for angles from $0^\circ-60^\circ$. For angles beyond $65^\circ$, the measurement results drop more quickly due to spill-over loss caused by the finite size of the radome.}
\label{fig:QO_measurement}
\end{figure*}

\section{Discussions and Conclusions}\label{sec:discussion}
{\color{black}
Until the recent developments in metamaterials, the antireflection structures developed for either optical or microwave applications were often limited to the use of isotropic dielectrics since the discovery of the antireflection phenomenon by Lord Rayleigh in the 19th century. However, most of the metamaterial based matching layers have similar structures to frequency selective surfaces, which often employ metallic strip or patch arrays. From the effective medium point of view, the metallic elements usually only increase the tangential effective permittivity due to the capacitive coupling between them. In this aspect, these metamaterial layers are not so different than conventional dielectric matching layers, which usually have a worse matching performance for the TE polarization due to the lack of control of the effective permeability. Although some metamaterial layers do achieve control of the permeability through the use of split-ring resonators (SRR), these resonant-based structures are often lossy and narrowband. In addition, the effective permittivity of the SRR structure is usually not of concern. To our best knowledge, for metamaterial based matching layers and absorbers, simultaneous control of the transversal and longitudinal permittivities and permeabilities (a total of 4 material parameters) has not been demonstrated.
In this regard, unlike previous reported metamaterial-based matching layers and absorbers, we demonstrated such a sophisticated control of all four materials parameters in the MEUML synthesis. The tangential and longitudinal permittivities, and permeabilities are judiciously tuned to achieve the required values. As a result, the MEUML is able to simultaneously match both TE and TM-polarized waves at oblique incident angles. We demonstrated this concept by synthesizing a physical MEUML which achieves a near perfect matching for the high-index RO3010 substrate ($\varepsilon_r=10.2)$ at $45^\circ$. The resulting reflection levels are around -30 dB for both polarizations.

Compared to our previous work \cite{7936430}, the MEUML is much more capable. In \cite{7936430}, the anisotropic matching layer is designed to match either the TE or the TM polarization, but not both; and the incident wave is limited to a particular orientation in the azimuthal plane ($y-z$ plane). On the other hand, the MEUML matches simultaneously both TE and TM polarizations, and there is no limit on the orientation of the incident plane wave. This enhanced capability comes at the cost of much stringent design constraints. In \cite{7936430}, the matching conditions are stipulated by two equations, but we have three materials parameters to satisfy the two constraints. For example, when matching for TE polarization, we use $\varepsilon_{yy}$, $\mu_{xx}$ and $\varepsilon_{zz}$ to satisfy those two equations. Thus, we have more degrees of freedom in the synthesis procedure. In comparison, the MEUML has 4 constraints (described by \eqref{Eq:impedance_relation} and \eqref{Eq:phase_relation}) that need to be satisfied by exactly 4 material parameter ($\varepsilon_{2t}$, $\mu_{2t}$, $\varepsilon_{2n}$, and $\mu_{2n}$); hence, there are no extra degrees of freedom. Furthermore, in \cite{7936430}, the structure we synthesized only needs to satisfy the material parameter values in one particular orientation. For example, we only need to satisfy $\varepsilon_{yy}$ but not $\varepsilon_{xx}$ because we limit the incident plane to be in the $y-z$ plane. For the MEUML, we need to satisfy the transversal $\varepsilon_{2t}$ parameter, which encompasses every orientation. Compared to \cite{7936430}, the synthesis of the MEUML is in fact much more challenging since we have no extra degrees of freedom, and the unit cell has to work for both polarizations and all orientations. This calls for a completely different approach and unit cell design.

Despite the fact that we are quite constraint in the MEUML synthesis, we still have the liberty to relax some constraints such that extreme or negative material parameters are not required. Thus, we do not have to rely on resonances, which otherwise would make the structure sensitive, narrow-band and lossy, to achieve those extreme material parameters. In the simulations, we showed that the MEUML is low-loss and broadband, which makes it very attractive for practical applications. 
We demonstrated the practicality of the MEUML by designing a novel sandwich type radome for microwave applications. We were able to show that a $\lambda_0/3$ thick radome can maintain reflection levels below -14 dB over the entire angular range of $0^\circ-85^\circ$ for both polarizations, and over a wide bandwidth. To our best knowledge, this is the widest angular range that has been reported in the literature for such a compact radome. 
The transmission of the fabricated radome is characterized by a wideband far-field measurement and a narrowband quasi-optical measurement. From the far-field measurements, we can conclude that the radome is indeed wideband and wide angle albeit the measured transmissions is less accurate. More accurate transmissions were obtained from the narrowband quasi-optical measurement. We observed an excellent agreement between the simulated and measured results for angles between $0^\circ-65^\circ$. For angles beyond $65^\circ$, the measured results are less accurate due to spill-over loss. Even thought the quasi-optical measurement was conducted only at 10~GHz, with the combination of the wideband far-field measurements, the radome is indeed working as it was proposed.

In summary, we have proposed a novel matching theory based on magneto-electric uniaxial metamaterials. Remarkably we have proposed a novel, yet simple, metamaterial structure that can synthesize the necessary  uniaxial parameters. The concept has been verified both in simulation as well as in measurement. Although the simulations and measurements were performed in the microwave regime, the theory is general and can be applied in the optical and terahertz regimes. With the help of a provided synthesis procedure and a proposed parameter retrieval technique, we can tune the required uniaxial parameters  precisely. We believe the MEUML concept not only opens new routes for impedance matching applications, but also opens new possibilities for designing more sophisticated metamaterials that provide provide advanced functionalities.}

\section*{Acknowledgment}
The authors would like to thank Rogers Corporations \cite{RogersCorp} for providing free substrate samples. Financial support has been provided by the Natural Sciences and Engineering Research Council of Canada (NSERC).

{\color{black} 

\appendix
\section{Derivation of the MEUML material parameter tensors for perfect matching}\label{sec:MEUML_parameter_derivation}
The total reflection in air is given by \eqref{Eq:appendix_total_r}
\begin{equation}
r^\mathrm{TM/TE}=\frac{r_{12}^\mathrm{TM/TE}+r_{23}^\mathrm{TM/TE}e^{-i2\phi^\mathrm{TM/TE}}}{1-r_{12}^\mathrm{TM/TE}r_{23}^\mathrm{TM/TE}e^{-i2\phi^\mathrm{TM/TE}}}
\label{Eq:appendix_total_r}
\end{equation}
where $r_{ij}^\mathrm{TM/TE}$ is the Fresnel reflection coefficient at the $i$, $j$ media interface with the incidence from medium $i$; the phase $\phi^\mathrm{TM/TE}$ is the total phase delay accumulated in the uniaxial layer along the normal of the layer surface (z-axis). 
Specifically,
\begin{equation}
r_{ij}^\mathrm{TM/TE}=\frac{Z_j^\mathrm{TM/TE}-Z_i^\mathrm{TM/TE}}{Z_j^\mathrm{TM/TE}+Z_i^\mathrm{TM/TE}}
\label{Eq:r_ij}
\end{equation}
\begin{equation}
\phi^\mathrm{TM/TE}=k_{2z}^\mathrm{TM/TE}d
\label{Eq:phi}
\end{equation}
where $Z_i^\mathrm{TM/TE}$ is the wave impedance for TM or TE polarization in medium $i$, $k_{2z}^\mathrm{TM/TE}$ is the wave number along the z-axis, and $d$ is the layer thickness. The wave impedances and the wave numbers can be further  expressed as
\begin{equation}
Z_1^\mathrm{TM}=\eta_0\cos\theta_1
\label{Eq:Z1_TM}
\end{equation}
\begin{equation}
Z_2^\mathrm{TM}=\eta_0\sqrt{\frac{\mu_{2t}}{\varepsilon_{2t}}-\sin^2\theta_1\frac{1}{\varepsilon_{2t}\varepsilon_{2n}}}
\label{Eq:Z2_TM}
\end{equation}
\begin{equation}
Z_3^\mathrm{TM}=\frac{\eta_0\sqrt{1-\frac{\sin^2\theta_1}{\varepsilon_3}}}{\sqrt{\varepsilon_3}}
\label{Eq:Z3_TM}
\end{equation}
\begin{equation}
Z_1^\mathrm{TE}=\frac{\eta_0}{\cos\theta_1}
\label{Eq:Z1_TE}
\end{equation}
\begin{equation}
Z_2^\mathrm{TE}=\frac{\eta_0}{\sqrt{\frac{\varepsilon_{2t}}{\mu_{2t}}-\sin^2\theta_1\frac{1}{\mu_{2t}\mu_{2n}}}}
\label{Eq:Z2_TE}
\end{equation}
\begin{equation}
Z_3^\mathrm{TE}=\frac{\eta_0}{\sqrt{\varepsilon_3}\sqrt{1-\frac{\sin^2\theta_1}{\varepsilon_3}}}
\label{Eq:Z3_TE}
\end{equation}
\begin{equation}
k_{2z}^\mathrm{TM}=k_0\sqrt{\mu_{2t}\varepsilon_{2t}-\sin^2\theta_1 \frac{\varepsilon_{2t}}{\varepsilon_{2n}}}=k_0n_{2t}^{\mathrm{TM}}
\label{Eq:k2z_TM}
\end{equation}
\begin{equation}
k_{2z}^\mathrm{TE}=k_0\sqrt{\mu_{2t}\varepsilon_{2t}-\sin^2\theta_1 \frac{\mu_{2t}}{\mu_{2n}}}=k_0n_{2t}^{\mathrm{TE}}
\label{Eq:k2z_TE}
\end{equation}
where $\eta_0$ is the free space intrinsic impedance and $k_0$ is the free space wave number. The required conditions to achieve zero reflection are \cite{7936430}
\begin{equation}
\left(Z_2^\mathrm{TM/TE}\right)^2=Z_1^\mathrm{TM/TE}Z_3^\mathrm{TM/TE}
\label{Eq:impedance_relation}
\end{equation}
\begin{equation}
k_{2z}^\mathrm{TM/TE}=\frac{\pi}{2d}
\label{Eq:phase_relation}
\end{equation}

By substituting \eqref{Eq:Z1_TM}--\eqref{Eq:Z3_TE} into \eqref{Eq:impedance_relation} and \eqref{Eq:k2z_TM}, \eqref{Eq:k2z_TE} into \eqref{Eq:phase_relation}, we have a total of four equations with four unknown material parameters. Hence, the tangential and longitudinal permittivities and permeabilities can be solved exactly:
\begin{equation}
\varepsilon_{2t}=\frac{\lambda_0}{4d}\frac{\sqrt{\varepsilon_3}}{\sqrt{\cos\theta_1\sqrt{\varepsilon_3-\sin^2\theta_1}}}
\label{Eq:Supp_eps_2t}
\end{equation}
\begin{equation}
\mu_{2t}=\frac{\lambda_0}{4d}\frac{1}{\sqrt{\cos\theta_1\sqrt{\varepsilon_3-\sin^2\theta_1}}}
\label{Eq:Supp_mu_2t}
\end{equation}
\begin{equation}
\varepsilon_{2n}=\frac{4d}{\lambda_0}\frac{\sqrt{\varepsilon_3}\sin^2\theta_1\sqrt{\cos\theta_1\sqrt{\varepsilon_3-\sin^2\theta_1}}}{\sqrt{\varepsilon_3}-\cos\theta_1\sqrt{\varepsilon_3-\sin^2\theta_1}}
\label{Eq:Supp_eps_2n}
\end{equation}
\begin{equation}
\mu_{2n}=\frac{4d}{\lambda_0}\frac{\sin^2\theta_1\sqrt{\cos\theta_1\sqrt{\varepsilon_3-\sin^2\theta_1}}}{\sqrt{\varepsilon_3}-\cos\theta_1\sqrt{\varepsilon_3-\sin^2\theta_1}}
\label{Eq:Supp_mu_2n}
\end{equation}
where $\lambda_0$ is the free space wavelength. Notice that 
\begin{equation}
\frac{\varepsilon_{2t}}{\mu_{2t}}=\frac{\varepsilon_{2n}}{\mu_{2n}}=\sqrt{\varepsilon_3}
\label{Eq::Supp_parameter_relation}
\end{equation}
The general trends of the MEUML parameters vs. $\theta_1$, $\varepsilon_3$, and $d$ are shown in Fig.~\ref{fig:required_layer_parameter} and Table.~\ref{tab:parameter_trend}

\begin{figure}[htbp!]
    \centering
  \subfloat[]{%
       \includegraphics[width=0.6\linewidth]{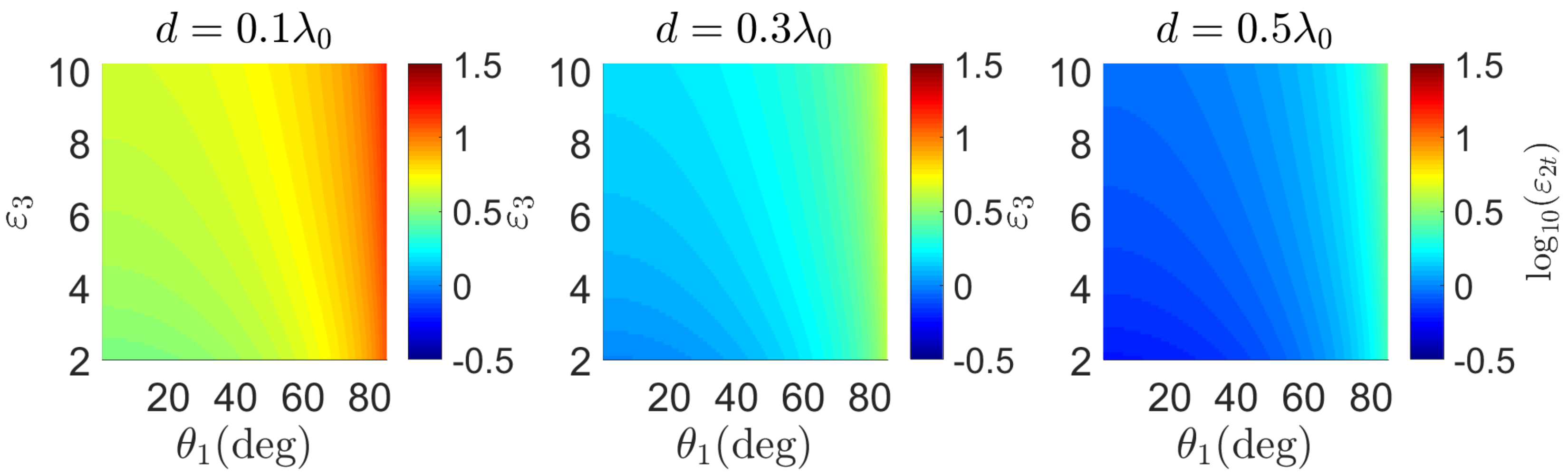}}
    \label{eps_2t}\\
  \subfloat[]{%
        \includegraphics[width=0.6\linewidth]{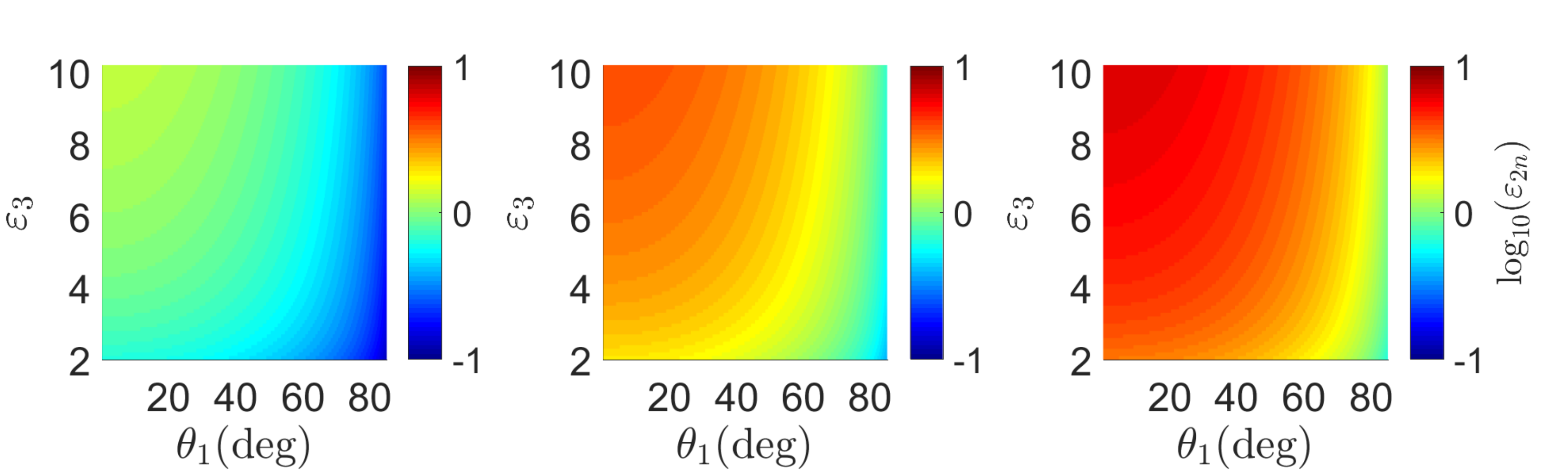}}
    \label{eps_2n}\\
  \subfloat[]{%
        \includegraphics[width=0.6\linewidth]{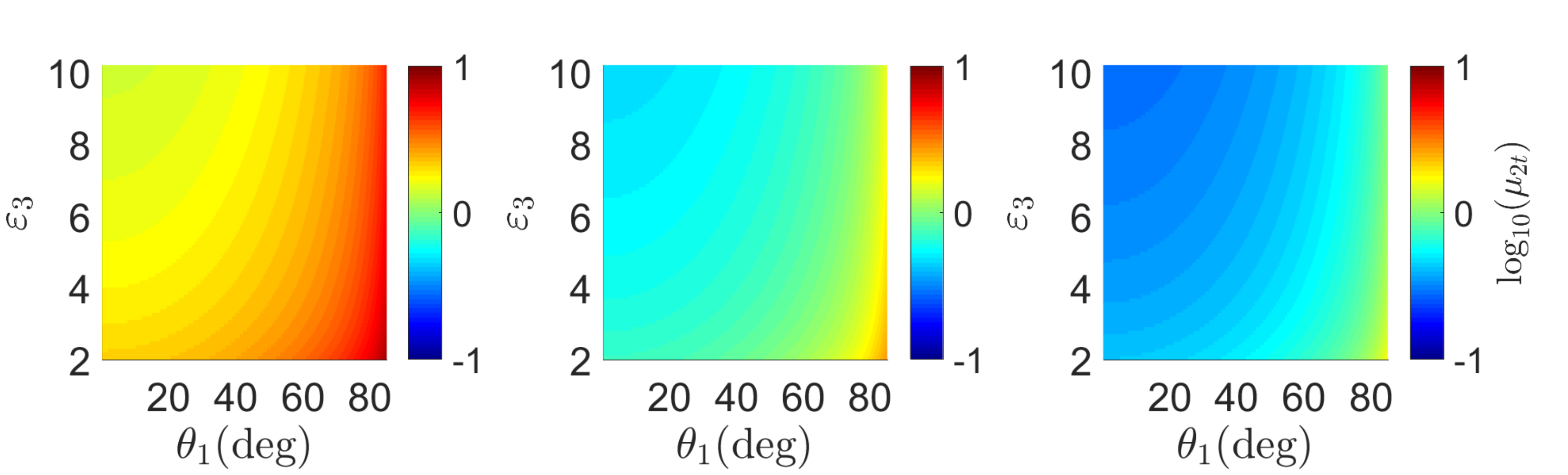}}
     \label{mu_2t} 
	\subfloat[]{%
        \includegraphics[width=0.6\linewidth]{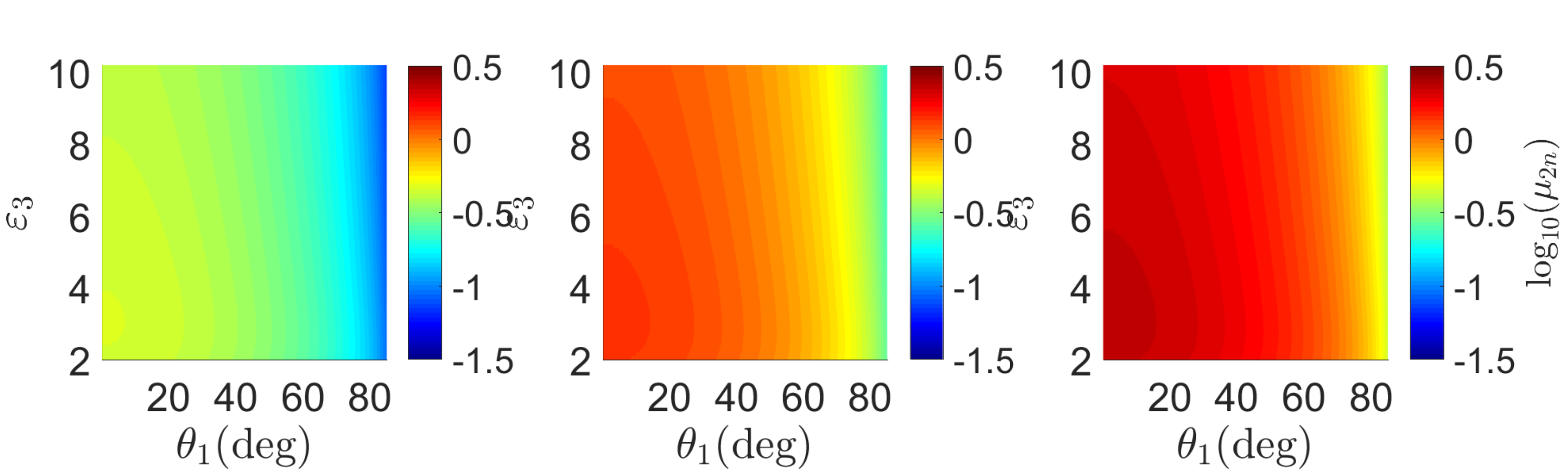}}
     \label{mu_2n} 
  \caption{Required (a) $\varepsilon_{2t}$, (b) $\varepsilon_{2n}$, (c) $\mu_{2t}$, and (d) $\mu_{2n}$ for perfect matching are plotted as functions of $\varepsilon_3$ and incident angle $\theta_1$ for three different MEUML thicknesses. The values for the material parameters are plotted in logarithmic scale for better visualization.}
  \label{fig:required_layer_parameter} 
\end{figure}

\begin{table}[htbp!]
\renewcommand{\arraystretch}{1.3}
\caption{General trends of the layer parameters vs. increasing $\theta_1$, $\varepsilon_3$, and $d$}
\label{tab:parameter_trend}
\centering
\begin{tabular}{|c||c|c|c|c|}
\hline
			 & $\varepsilon_{2t}$ & $\varepsilon_{2n}$ & $\mu_{2t}$ & $\mu_{2n}$\\
			\hline
			$\theta_1 \uparrow$ & $\uparrow$ & $\downarrow$ & $\uparrow$ & $\downarrow$\\
			\hline
			$\varepsilon_3 \uparrow$ & $\uparrow$ & $\uparrow$ & $\downarrow$ & $\downarrow$\\
			\hline
			$d \uparrow$ & $\downarrow$ & $\uparrow$ & $\downarrow$ & $\uparrow$\\
\hline
\end{tabular}
\end{table}

\section{Physical MEUML synthesis procedure}\label{sec:MEUML_synthesis}
To design the MEUML to match to a particular substrate at a particular incident angle, the following iterative synthesis procedure is proposed:
\begin{enumerate}
	\item Define the desired incident angle $\theta_1$ to achieve perfect matching for the substrate with $\varepsilon_3$. 
	\item Define an initial layer thickness $d$ and calculate the required layer parameter values from (4)--(7) in the manuscript.
	\item Based on the calculated parameter values, propose an initial structure and extract the layer parameters.
	\item Based on the extracted parameter values, one has the following options:
	\begin{itemize}
		\item If the extracted values are close to the desired ones, one can optimize the structure geometries such that the extracted values approach the desired ones.
		\item If the extracted values are not close to the desired ones, one can either revise the layer thickness in step (2) or the proposed structure in step (3) and repeat the subsequent iterative procedure.
	\end{itemize}
\end{enumerate}

For a proof of concept, matching a Rogers RO3010 substrate ($\varepsilon_r=10.2$) at $45^\circ$ is demonstrated. Using (4)--(7) in the manuscript, the required material parameters as a function of $\theta_1$ for a few substrate layer thicknesses are plotted in Fig.~\ref{fig:layer_parameters_45deg}. 
To determine a good initial layer thickness, one should consider the loss, sensitivity and ease of synthesis of the layer. 
In terms of the synthesis, it is generally easier to realize material parameters with less contrast, i.e., the longitudinal and tangential permittivities or permeabilities do not differ too much from each other. 
In terms of loss, notice that a thinner layer corresponds to a higher $\varepsilon_{2t}$. Since the effective permittivity of typical metamaterials can be characterized by the Drudes's model, the electrical response of such materials can be described by the Kramer-Kronig relation, which states that the real and imaginary parts of  of the permittivity are not independent from each other. A material with a larger $\Re\{\varepsilon\}$ is often associated with a more negative $\Im\{\varepsilon\}$, suggesting the material is more dispersive and dissipates more energy. In this aspect, a thinner layer can be more lossy. 
In terms of sensitivity, the structure should not operate near its resonance. From the authors' previous investigations, it was found that realizing a layer with a strong para-magnetic property (large $\mu$) relies on the resonance of the metamaterial structure. Such a layer is not only more lossy, but also more sensitive than a layer with a weak para-magnetic property ($\mu\approx 1$) or a diamagnetic property ($\mu<1$).
Bearing those considerations in mind, in order to synthesize a low-loss and insensitive matching layer, we conclude that the required permittivities should not be too high and the permeabilities should be close to or smaller than unity. From Fig.~\ref{fig:layer_parameters_45deg}, a layer thickness of 5mm is a good starting point in the iterative procedure. The final thickness of the layer is revised to 4.75mm, which is an easier thickness to realize in fabrication.
\begin{figure}[htbp!]
\centering
\includegraphics[width=0.55\columnwidth]{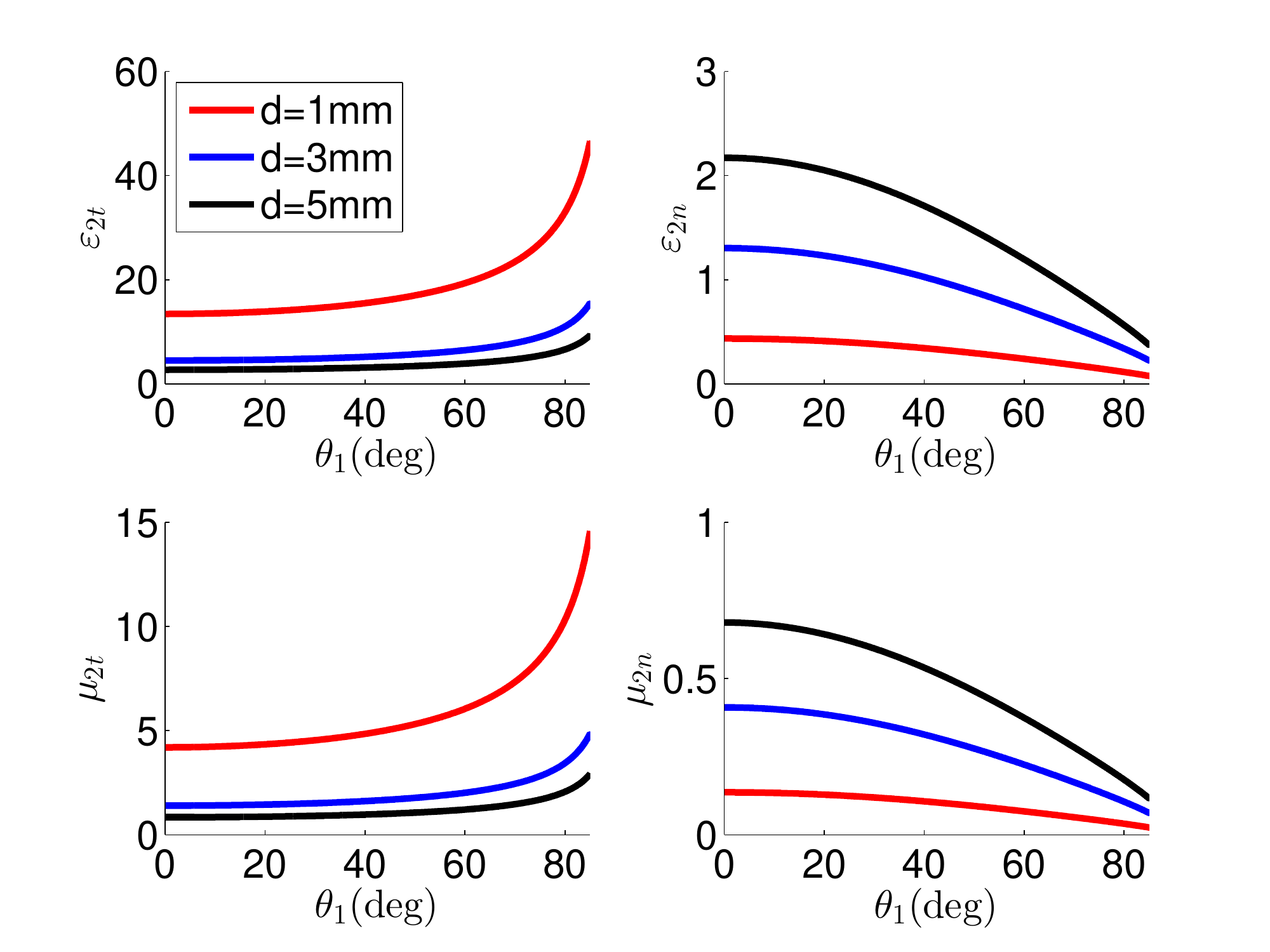}
\caption{The required material parameters to match to a high-index substrate with $\varepsilon_3=10.2$ are plotted against $\theta_1$ for three layer thicknesses.}
\label{fig:layer_parameters_45deg}
\end{figure}

\section{MEUML numerical simulation and parameter extraction}\label{sec:MEUML_parameter_extraction}
Full-wave electromagnetic simulation of the MEUML unit cell is performed in ANSYS HFSS commercial software. Two Floquet ports are used with periodic boundary conditions. The host medium of the MEUML is a Rogers 5870 substrate with $\varepsilon_r=2.33$. Since the 4.75mm thickness is not readily available, it is realized by bonding a standard 3.175mm thick substrate to two 0.787mm ones. The bonding film between the substrates is the Rogers 3001 bonding film with a thickness of 0.0381mm and a permittivity of 2.28. The thin bonding films are included in the simulation. The copper and dielectric losses are also accounted for in the simulation. From the unit cell simulations, we can obtain the scattering parameters of the MEUML and they are used in the parameter extraction procedure, which is adapted from the method presented in the Appendix of \cite{7936430}. It takes into account of both the angle of incidence and the high-index substrate effect. 

$n_{2t}^{\mathrm{TM}}$ and $n_{2t}^{\mathrm{TE}}$ of the MEUML in \eqref{Eq:k2z_TM} and \eqref{Eq:k2z_TE} can be related to the scattering parameters through \eqref{Eq:nt_extraction}

\begin{equation}
n_{2t}^\mathrm{TM,TE}=\frac{1}{k_0 d}\cos^{-1}\left(\frac{1-S_{11} S_{22}+S_{21} S_{12}}{S_{21}+S_{12}}\right)^\mathrm{TM,TE}
\label{Eq:nt_extraction}
\end{equation}
where $k_0$ is the free space wave number and $d$ is the MEUML thickness. The wave impedances of the MEUML in \eqref{Eq:Z2_TM} and \eqref{Eq:Z2_TE} can be related to the S-parameters through \eqref{Eq:Z2normTM}-\eqref{Eq:Z2TM_extraction}

\begin{equation}
Z_{2,norm}^\mathrm{TM}=\frac{Z_2^\mathrm{TM}}{Z_1^\mathrm{TM}}=\frac{Z_2^\mathrm{TM}}{\eta_0\cos\theta_1}
\label{Eq:Z2normTM}
\end{equation}

\begin{equation}
Z_{2,norm}^\mathrm{TE}=\frac{Z_2^\mathrm{TE}}{Z_1^\mathrm{TE}}=\frac{Z_2^\mathrm{TE}\cos\theta_1}{\eta_0}
\label{Eq:Z2normTE}
\end{equation}

\begin{equation}
Z_{2,norm}^\mathrm{TM}=\sqrt{\frac{\sqrt{\varepsilon_3-\sin^2\theta_1}}{\varepsilon_3}}\sqrt{\frac{(1+S_{11})(1+S_{22})-S_{21}S_{12}}{(1-S_{11})(1-S_{22})-S_{21}S_{12}}}
\label{Eq:Z2TM_extraction}
\end{equation}

\begin{equation}
Z_{2, norm}^\mathrm{TE}=\sqrt{\frac{\cos\theta_1}{\sqrt{\varepsilon_3-\sin^2\theta_1}}}\sqrt{\frac{(1+S_{11})(1+S_{22})-S_{21}S_{12}}{(1-S_{11})(1-S_{22})-S_{21}S_{12}}}
\label{Eq:Z2TE_extraction}
\end{equation}

Using the extracted indices and wave impedances, $\varepsilon_{2t}$ and $\mu_{2t}$ can be obtained as in \eqref{Eq:appendix_eps_2t} and \eqref{Eq:appendix_mu_2t}. Once $\varepsilon_{2t}$ and $\mu_{2t}$ are obtained, they can be substituted into \eqref{Eq:k2z_TM} and \eqref{Eq:k2z_TE} to obtain $\varepsilon_{2n}$ and $\mu_{2n}$ in \eqref{Eq:appendix_eps_2n} and \eqref{Eq:appendix_mu_2n}. 

\begin{equation}
\varepsilon_{2t}=\frac{n_{2t}^\mathrm{TM}}{\cos\theta_1 Z_{2,norm}^\mathrm{TM}}
\label{Eq:appendix_eps_2t}
\end{equation}

\begin{equation}
\mu_{2t}=\frac{n_{2t}^\mathrm{TE} Z_{2,norm}^\mathrm{TE}}{\cos\theta_1 }
\label{Eq:appendix_mu_2t}
\end{equation}

\begin{equation}
\varepsilon_{2n}=\frac{\varepsilon_{2t}\sin^2\theta_1}{\varepsilon_{2t}\mu_{2t}-\left(n_{2t}^\mathrm{TM}\right)^2}
\label{Eq:appendix_eps_2n}
\end{equation}

\begin{equation}
\mu_{2n}=\frac{\mu_{2t}\sin^2\theta_1}{\varepsilon_{2t}\mu_{2t}-\left(n_{2t}^\mathrm{TE}\right)^2}
\label{Eq:appendix_mu_2n}
\end{equation}
Notice that the wave impedances and wave numbers are independent from the longitudinal parameters at normal incidence. Thus, in Fig.~\ref{fig:extracted_layer_parameters}, the longitudinal parameters of the MEUML cannot be extracted at normal incidence. 

\section{TMM analysis of the MEUML radome}\label{sec:TMM analysis of the MEUML radome}
The performance of the radome is analyzed by the transfer matrix method and the transfer matrix is given by \eqref{eqn:transfer_matrix}
\begin{equation}
  \begin{bmatrix}
    a_0 \\
    b_0 
  \end{bmatrix}
	=\frac{\mathcal{T}_{01}\mathcal{P}_{1}\mathcal{T}_{12}\mathcal{P}_{2}\mathcal{T}_{21}\mathcal{P}_{1}\mathcal{T}_{10}}{t_{01}t_{12}t_{21}t_{10}}
		 \begin{bmatrix}
    a_4 \\
    b_4 
  \end{bmatrix}	
	=\left[\mathcal{M}\right]
	 \begin{bmatrix}
    a_4 \\
    b_4 
  \end{bmatrix}	
\label{eqn:transfer_matrix}
\end{equation}
where $a_i$ and $b_i$ are the wave amplitudes in medium $i$. $\mathcal{T}_{ij}$ and $\mathcal{P}_{k}$ are defined as
\begin{equation}
\mathcal{T}_{ij}=
  \begin{bmatrix}
    1 & r_{ij}\\
    r_{ij} & 1
  \end{bmatrix}
\label{eqn:T_ij}
\end{equation}
\begin{equation}
\mathcal{P}_{k}=
  \begin{bmatrix}
    e^{j\phi_k} & 0\\
    0 & e^{-j\phi_k}
  \end{bmatrix}
\label{eqn:P_k}
\end{equation}
where $r_{ij}$ and $t_{ij}$ are the Fresnel coefficients at each layer interface, and $\phi_k$ is the phase delay in each layer. 
The terms in \eqref{eqn:transfer_matrix} can be grouped as in \eqref{eqn:appendix_transfer_matrix_grouped} to gain insight in the total reflection of the sandwich structure. $\frac{\mathcal{T}_{01}\mathcal{P}_{1}\mathcal{T}_{12}}{t_{01}t_{12}}$ or $ \frac{\mathcal{T}_{21}\mathcal{P}_{1}\mathcal{T}_{10}}{t_{21}t_{10}}$ is in fact the transfer matrix for calculating the reflection and the transmission of an MEUML, sandwiched between a semi-infinite air space and a semi-infinite substrate. $t_{02}$ is the total transmission from the semi-infinite air region to the semi-infinite substrate; $t_{20}$ is the total transmission from the semi-infinite substrate to the semi-infinite air region; $r_{02}$/$r_{20}$ is the total reflection with the incidence from the semi-infinite air/substrate. Thus, $r_{02}$ and $t_{02}$ are equivalent to $r$ and $t$ in Fig.~5 of the manuscript.

\begin{multline}
	 \begin{bmatrix}
    \mathcal{M}_{11} & 	\mathcal{M}_{12} \\
    \mathcal{M}_{21} &  \mathcal{M}_{22}
   \end{bmatrix}
=\frac{\mathcal{T}_{01}\mathcal{P}_{1}\mathcal{T}_{12}}{t_{01}t_{12}} \mathcal{P}_{2} \frac{\mathcal{T}_{21}\mathcal{P}_{1}\mathcal{T}_{10}}{t_{21}t_{10}}
	= \begin{bmatrix}
    \frac{1}{t_{02}} & -\frac{r_{20}}{t_{02}} \\
    \frac{r_{02}}{t_{02}} &  \frac{t_{02}t_{20}-r_{02}r_{20}}{t_{02}}
   \end{bmatrix}
	 \begin{bmatrix}
    e^{i\phi_2} & 	0 \\
    0 					&  	e^{-i\phi_2}
   \end{bmatrix}	
	 \begin{bmatrix}
    \frac{1}{t_{20}} & -\frac{r_{20}}{t_{20}} \\
    \frac{r_{02}}{t_{20}} &  \frac{t_{02}t_{20}-r_{02}r_{20}}{t_{20}}
   \end{bmatrix}	
\label{eqn:appendix_transfer_matrix_grouped}
\end{multline}
The total reflection of the sandwich structure is given by \eqref{eq:appendix_sandwich_reflection} 
\begin{equation}
r_{sandwich}=\frac{\mathcal{M}_{21}}{\mathcal{M}_{11}}=\frac{r_{02}e^{i\phi_2}-r_{02}e^{-i\phi_2}(t_{20}t_{02}-r_{02}r_{20})}{e^{i\phi_2}-e^{-i\phi_2}r_{02}r_{20}}
\label{eq:appendix_sandwich_reflection}
\end{equation}
Using \eqref{eq:sandwich_reflection}, we can calculate the maximum reflection within the angular range of $0^\circ-85^\circ$ as a function of the RO3010 substrate thickness. The results is plotted in Fig.~\ref{fig:Maximum_stack_reflection}.
\begin{figure}[!htbp]
\centering
\includegraphics[width=0.35\columnwidth]{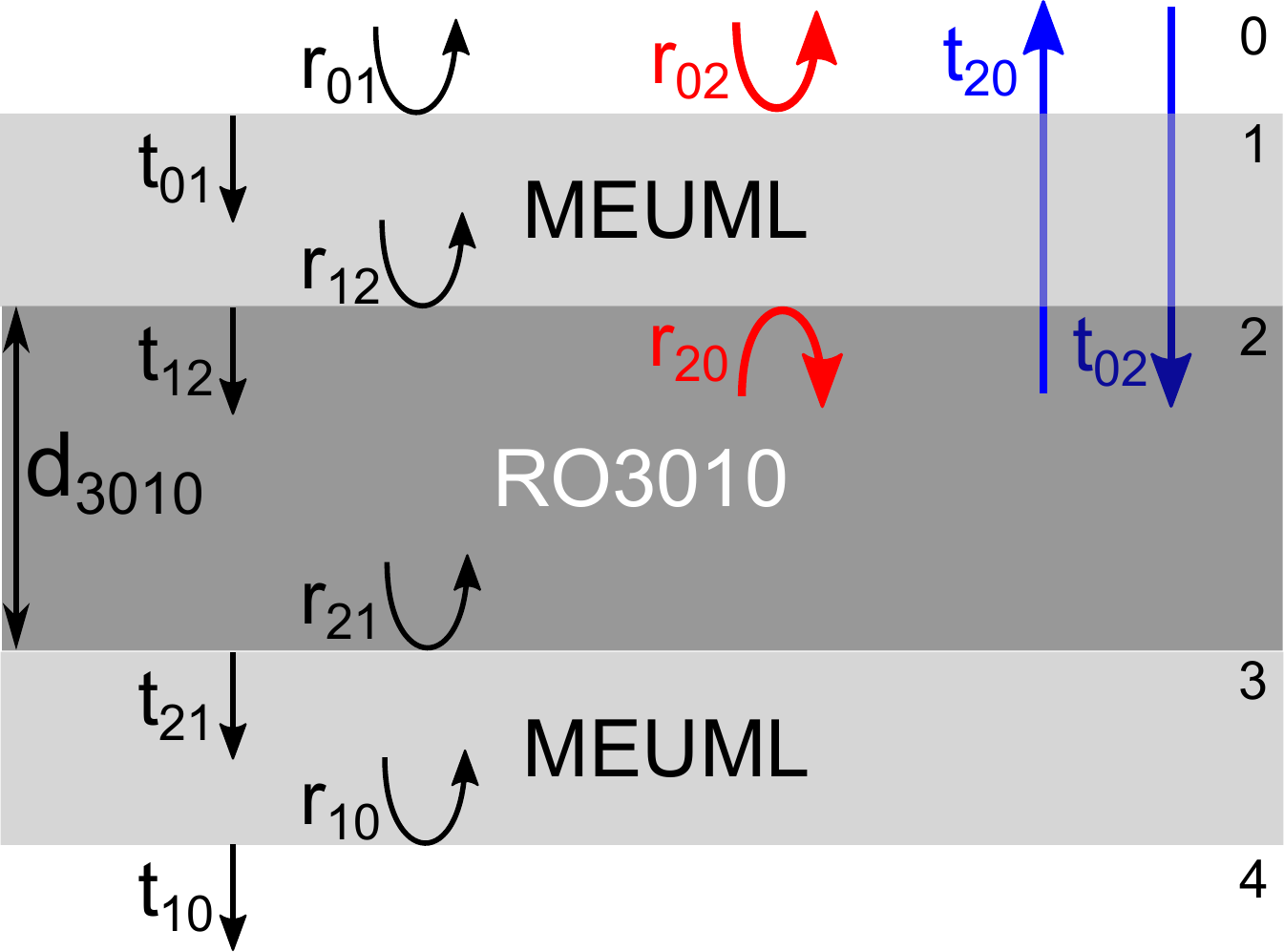}
\caption{The radome consists of a RO3010 substrate sandwiched by two MEUMLs. The transfer matrix method is used to analyze the total reflection of the sandwich.}
\label{fig:radome_layer_drawing}
\end{figure}
\begin{figure}[!htbp]
\centering
\includegraphics[width=0.5\columnwidth]{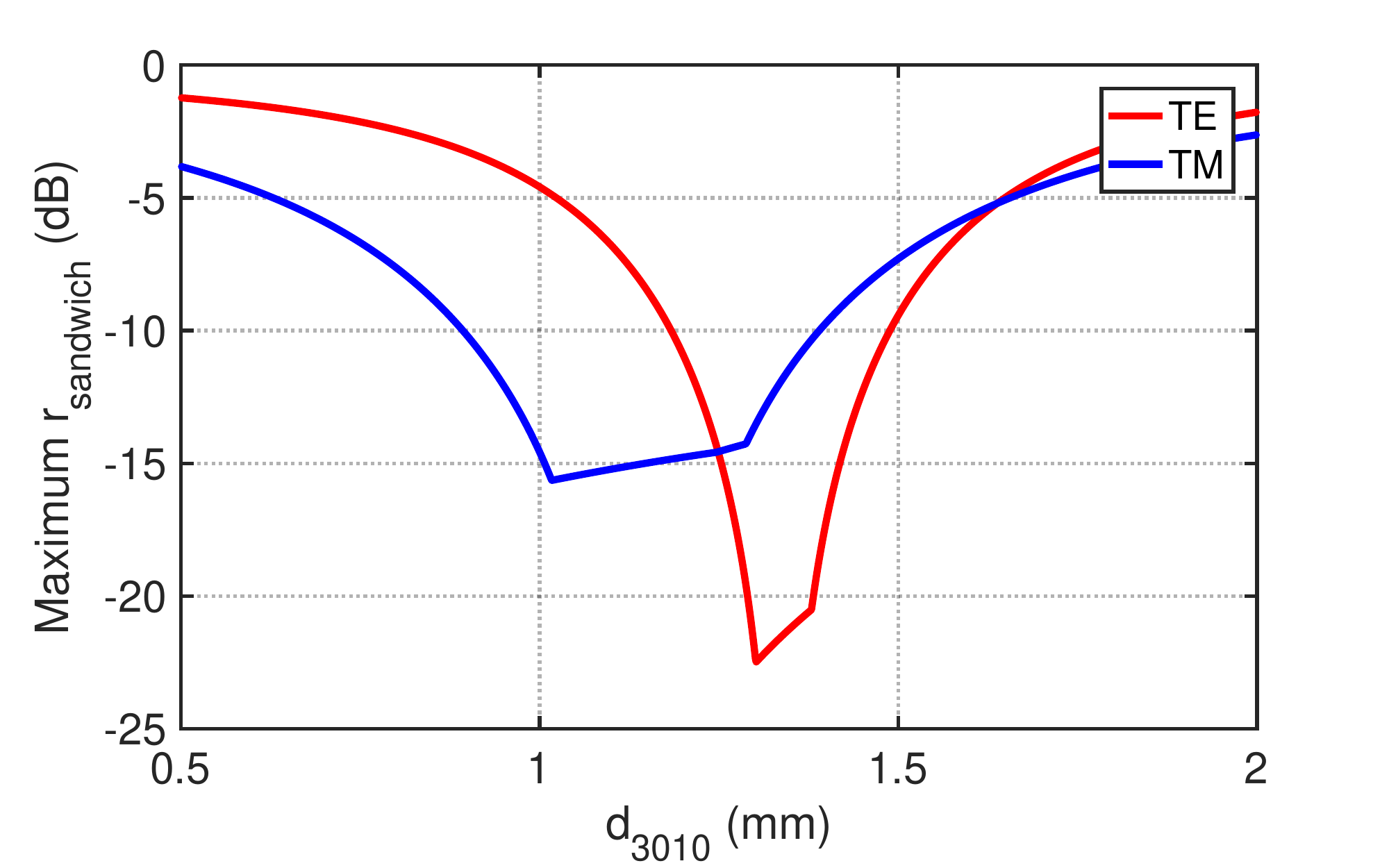}
\caption{{Maximum radome reflection as a function of RO3010 thickness}. Based on the transfer matrix method, the maximum reflection of the MEUML radome is calculated for the angular range of $0^\circ-85^\circ$.}
\label{fig:Maximum_stack_reflection}
\end{figure}

\section{MEUML radome fabrication and measurement}\label{sec:MEUML Fabrication and Measurement}
The expanded view of the MEUML radome is shown in Fig.~\ref{fig:FarField_measurement}(b). The MEUML consists of three layers: two 0.787mm and one 3.175mm Rogers RO5870 substrate. The copper rings are patterned on the 0.787mm substrates and the holes are drilled on the 3.175mm one. After the patterning and the drilling processes, the three substrates are bonded together with the Rogers 3001 bonding film to form the MEUML. The fabricated MEUML is cut in half and bonded to two sides of the 1.27mm thick Rogers 3010 substrate. The resulting structure is the MEUML radome. The patterning, drilling and alignment errors are better than 0.1mm.

The far-field measurement setup is shown in Fig.~\ref{fig:FarField_measurement}(a). A small slot antenna was placed behind the MEUML radome with a 1.5~cm thick foam as a spacer. Two gain measurements were conducted with and without the radome in front of the antenna. The transmission of the radome can be extracted from the difference of the gains. As the pedestal rotated, the transmission of the radome was measured against the incident angle. The transmitting pyramidal horn can be rotated such that the radiated electric field is polarized either horizontally or vertically. Thus, the transmission of the radome for a TE or a TM-polarized wave can be measured accordingly. The transmitting pyramid horn has a bandwidth of $8.2-12.4$~GHz, so it is difficult to use a single antenna receiver to cover the entire bandwidth. Three slot antennas were designed as shown in Fig.~\ref{fig:fabricated_slot_antenna} and each covers a different band. By stitching the results from each band, the radome can be characterized over the entire bandwidth.

\begin{figure}[htbp!]
\centering
\includegraphics[width=0.5\columnwidth]{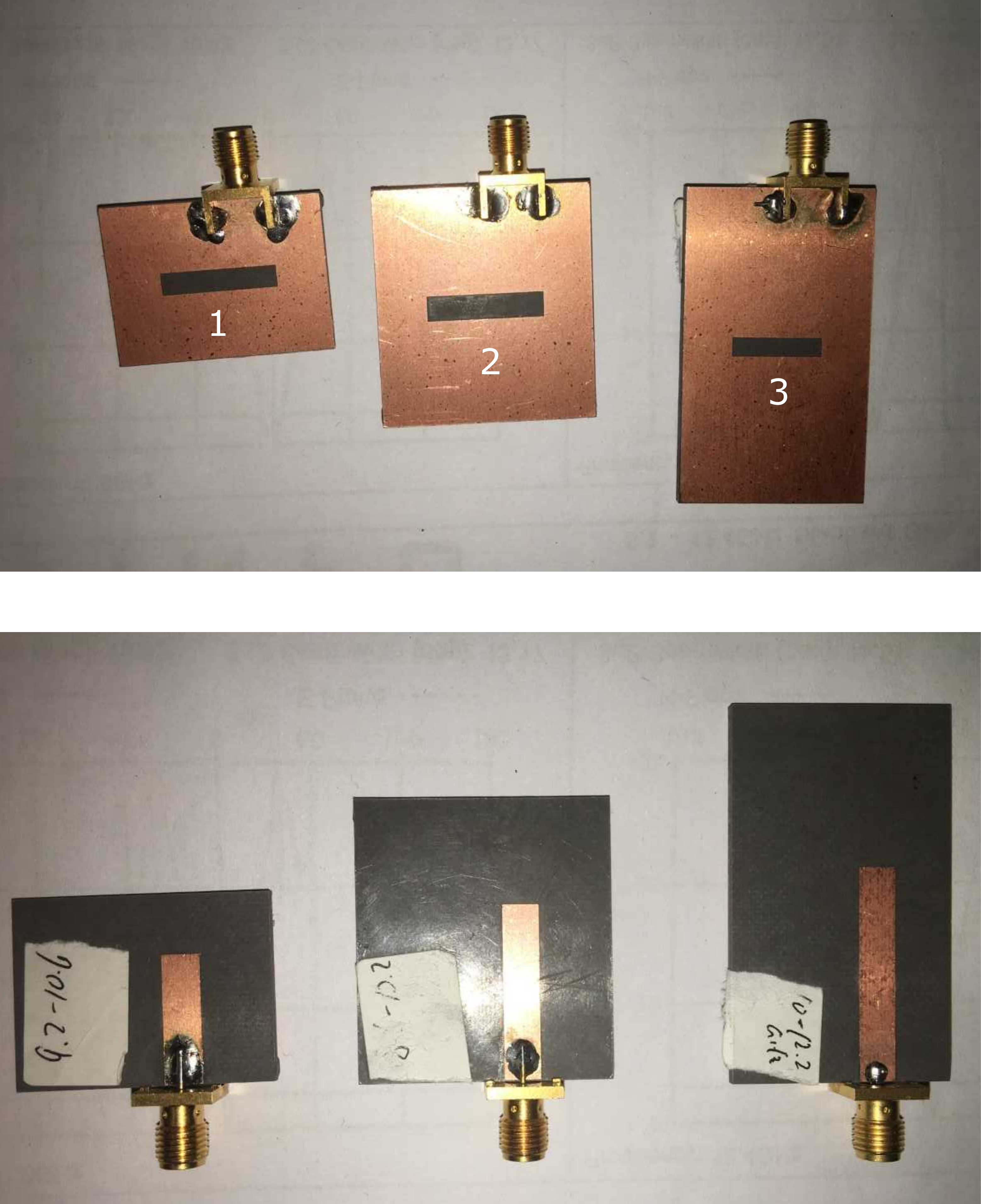}
\caption{{Fabricated receiving slot antennas for far-field measurements.} Since the transmitting horn has a wide bandwidth of $8.2-12.4$~GHz, three slot antennas were used to cover the entire bandwidth. Slot antennas 1, 2, and 3 have -10~dB reflections between $8.6-10.2$~GHz, $9.2-10.6$~GHz, and $10-12.2$~GHz, respectively.}
\label{fig:fabricated_slot_antenna}
\end{figure}

The quasi-optical setup is shown in Fig.~\ref{fig:QO_measurement}. It consists of two dual-polarized X-band horns and two Rexolite ($\varepsilon_r= 2.1$) lenses.The electromagnetic radiation from the horns is collimated by Rexolite lenses. A plane-wave incidence on the radome is emulated by placing it at the collimated Gaussian beam-waist. Since the collimating lens has a high numerical aperture, the location of the Gaussian beam-waist would shift with frequency. As a result, the quasi-optical setup is narrowband and the radome was only measured at the design frequency of 10~GHz. Since the horn used is dual-polarized, TE and TM transmissions of the radome can be characterized simultaneously as the radome rotates. The rotation of the MEUML is achieved through an electronically controlled rotational stage. The transmission of the radome is measured every $5^\circ$.

}
\bibliography{thesis}
\bibliographystyle{apsrev4-1}

%

\end{document}